\documentclass[]{aastex631}

\usepackage{float}
\usepackage{booktabs}
\usepackage{multirow}
\usepackage[figuresright]{rotating}
\usepackage{natbib}
\usepackage{amsmath}
\usepackage{subfigure}
\usepackage{color}
\usepackage[section]{placeins}
\usepackage{array}
\usepackage{soul}
\soulregister\cite7
\soulregister\ref7 
\soulregister\citep7 
\soulregister\citet7 
\soulregister\citealp7 

\begin{document}

\title{Photometric and Spectroscopic Analysis of V583 Lyrae, an Algol with a $g$-mode Pulsating Primary and Accretion Disk}

\author{Hui-Ting Zhang}
\affiliation{Yunnan Observatories, Chinese Academy of Sciences (CAS), 396 yangfangwang, Guandu District, Kunming, 650216, P.R. China}
\affiliation{Department of Astronomy, School of Physics and Astronomy, Yunnan University, Kunming 650091, P. R. China}
\affiliation{Key Laboratory of Astroparticle Physics of Yunnan Province, Yunnan University, Kunming 650091, P. R. China}

\author[0000-0002-5995-0794]{Sheng-Bang Qian}
\altaffiliation{E-mail: qsb@ynao.ac.cn}
\affiliation{Department of Astronomy, School of Physics and Astronomy, Yunnan University, Kunming 650091, P. R. China}
\affiliation{Key Laboratory of Astroparticle Physics of Yunnan Province, Yunnan University, Kunming 650091, P. R. China}

\author[0000-0001-9346-9876]{Wen-Ping Liao}
\affiliation{Yunnan Observatories, Chinese Academy of Sciences (CAS), 396 yangfangwang, Guandu District, Kunming, 650216, P.R. China}
\affiliation{University of Chinese Academy of Sciences, No.1 Yanqihu East Rd, Huairou District, Beijing, PR China, 101408}

\author{B. Soonthornthum}
\affiliation{National Astronomical Research Institute of Thailand (NARIT), 260 Don Kaew, Mae Rim Chiang Mai, Thailand, 50180}

\author{N. Sarotsakulchai}
\affiliation{National Astronomical Research Institute of Thailand (NARIT), 260 Don Kaew, Mae Rim Chiang Mai, Thailand, 50180}




\begin{abstract}

V583 Lyr is an extremely low mass ratio Algol-type binary with an orbital period of 11.2580 days. We determined an effective temperature of $T_{eff1}$ = 9000 $\pm$ 350 K from newly observed spectra, which might be an underestimate due to binary mass transfer. The binary mass ratio $q$ = 0.1 $\pm$ 0.004 and the orbital inclination $i$ = 85$^{\circ}$.5 are determined based on the assumption that the secondary fills its Roche lobe and rotates synchronously. The radial velocity curve is obtained from time series spectra, allowing for improved estimation of stellar masses and radii: $M_{1}$ = 3.56 $\pm$ 0.5 $M_{\odot}$, $R_{1}$ = 2.4 $\pm$ 0.2 $R_{\odot}$; and $M_{2}$ = 0.36 $\pm$ 0.02 $M_{\odot}$, $R_{2}$= 6.9 $\pm$ 0.4 $R_{\odot}$. The variations in the double-peaked $H_{\alpha}$ emission indicate the formation of a stable disk during mass transfer. V583 Lyr appears to be a post-mass-reversal system, according to the estimated mass transfer using O-C period analysis. Its orbital period is slowly increasing, from which the rate of mass accretion by the primary star is estimated to be $\dot{M_{1}}$ = 3.384 $\times10^{-8}$ $M_{\odot}\cdot yr^{-1}$. The pulsation analysis was conducted on the residuals of the light curve. The primary component was found to be a $g$-mode pulsating star with 26 frequencies extracted lower than 9 $d^{-1}$. The frequency groups and rotational splitting properties of the $g$-mode were studied in detail. This study provides compelling evidence for an accretion disk surrounding the $g$-mode pulsating primary. 

\end{abstract}

\keywords{binaries: eclipsing ---  binaries: spectroscopic ---  accretion disk ---   stars: pulsation }


\section{Introduction}\label{sec:intro}
The accretion structures formed are mainly determined by the relative separation of the components, the radius of the primary, and the mass ratio \citep{1975ApJ...198..383L}. In short-period Algols with $P$ $\leq$ 4.5 days, the gainer is larger and the gas stream from the inner Lagrangian point can impact the star directly. While if the gainer has a radius smaller than a certain specified size, relative to the binary separation, the infalling material has too much angular momentum for the stream to directly impact the gainer. The mass flow will miss the gainer and may collide with itself, changing its radial motion and instead feeding into a stable disk \citep{1975ApJ...198..383L}. This behavior usually occurs in long-period systems ($P$ $\geq$ 6 d). And there remains the case of a gainer with the radius between the minimum and disk radius, where a transient disk may exist. These are intermediate-period systems studied by \cite{1998ApJ...500L..17P,1999ApJS..123..537R} and references therein. An accretion disk may exist in $\beta$ Lyrae-type \citep{2022Galax..10...15M} and long-period Algol-type binary stars \citep{2010MNRAS.406.1071D}. The disk precession \citep{10.1093/mnras/stv347, 10.1093/mnras/282.2.597} may cause periodic flux variations as the disk area projected in the direction of the observer's line of sight changes, which may be a compelling explanation for the long period variation in some double period variables (DPVs, e.g. AU Mon \citealp{1980A&A....85..342L,10.1111/j.1365-2966.2010.17310.x})

V583 Lyr was recently analyzed by \cite{Zhang_2020} and found to be a semi-detached binary with an extremely low mass ratio of ($q$ = $M_{2}/M_{1}$  = 0.096). \cite{2019A&A...630A.106G} suggested that V583 Lyr may be the type of $\delta$ Sct or tidal oscillations with frequencies larger than 5 $d^{-1}$ and effective temperature \citep{1973A&A....23..221B, 2015MNRAS.452.3073B}.  However, they did not look for regular frequency spacing or any kind of pattern that would indicate a $g$-mode $\gamma$ Dor star  \citep{1999PASP..111..840K, 2016A&A...593A.120V} or the slowly pulsating B (SPB, \citealp{1991A&A...246..453W}) stars. \cite{2015AJ....149..197Z} found evidence supporting the possibility of a third body with a 50-year period, based on variations in previous photographic data. And \cite{2016ApJ...829...23D}'s statistical search for stellar flares reported 191 high levels of flare activity that are thought to be signals of light changes from binary eclipses.

V583 Lyr is an Algol system with an extremely low mass ratio and an orbital period of 11.2580 days. The strong double-peaked $H_{\alpha}$ emission is present in all observed spectra, one of the more prominent observational indicators of a permanent accretion disk. The mass transfer or accretion in V583  Lyr can be discovered from a detailed analysis of the available observations. Furthermore, V583 Lyr exhibits oscillations in all the photometric observations. We use the methods of asteroseismology, which involves the analysis of stellar pulsations to study stellar structure and evolution \citep{2010aste.book.....A}, to provide a valuable supplement to the examination of V583 Lyr. 

This paper presents the photometric and spectral observations and data reductions on V583 Lyr in section \ref{sec:data}. From the spectroscopy study in section \ref{sec:Temperature}, the preliminary effective temperature of the hotter component is estimated. In section \ref{sec:photo}, the light curve synthesis was performed in conjunction with new data from \emph{TESS} and a reanalysis of the \emph{Kepler} data. To obtain the mass of the binary components, radial velocities are measured through spectroscopy in section \ref{sec:RV}. In section \ref{sec:H profiles}, we present a thorough spectroscopic analysis, highlighting the double-peaked $H_{\alpha}$ emission lines and the accretion disk. In section \ref{sec:Pulsation}, we performed an asteroseismic analysis of the photometry for V583 Lyr, examining the frequency groups and $g$-mode splitting to determine the pulsation type and rotation of the gainer. Finally, in section \ref{sec:Conclusion}, we provide the conclusions and discussions of our work.

\section{Observation and data reduction} \label{sec:data}

\subsection{Photometric Data and Resampling\label{subsec:photometry}}
V583 Lyr was cataloged in \emph{Kepler} Eclipsing Binary Stars \citep{2011AJ....141...83P}  for photometric data by \emph{Kepler} \citep{2010ApJ...713L..79K}. The time series photometry data from \emph{Kepler} between 2009 and 2013 for long-cadence (LC, 1800 seconds) light curves, and the short-cadence (SC, 60 seconds) data were observed for almost 3 cycles. Data are available at MAST \footnote{\url{https://archive.stsci.edu/kepler/data_search/search.php}} in both SC and LC formats. We downloaded all available \emph{TESS} (Transiting Exoplanet Survey Satellite) and \emph{Kepler} observations of V583 Lyr using the $\mathtt{lightkurve}$ package \citep{2018ascl.soft12013L}. New observations from \emph{TESS} were obtained in sectors 14, 40, 41, 53, and 54. The photometric band range for \emph{TESS} (600-1000 $nm$) aiming for a photometric accuracy of 50 ppm on stars \citep{2015JATIS...1a4003R}. The \emph{TESS} and \emph{Kepler} data points are phase-folded and resampled using the method suggested by \cite{Zhang_2020}.

\subsection{Spectroscopy data and reductions} \label{subsec:spectras}

The double-peaked $H_{\alpha}$ profile is observed by \emph{LAMOST} (The Large Sky Area Multi-Object Fiber Spectroscopic Telescope, \citealp{2022yCat.5156....0L}) Low-Resolution Spectroscopic Survey (LRS) in June 2017, which suggests the presence of an accretion disk. So we conducted more optical spectroscopic observations of V583 Lyr at different eclipsing phases with Beijing Faint Object Spectrograph and Camera (\emph{BFOSC}, \citealp{2016PASP..128k5005F}) of Xinglong 2.16 m telescope for direct imaging and low-resolution (R $\sim$ 500–2000) spectroscopy. Five spectra were obtained from February to May of 2023 with a spectral region of 387–676 nm. The \emph{BFOSC} observations are listed in Table \ref{tab:spec}. All the observations and \emph{LAMOST} LRS spectra (at phase of 0.66) cover the orbital circle roughly well. All spectra used in this work are normalized to the continuum approximated by spline functions, and the radial velocities (RVs) are heliocentric. The standard $\mathtt{IRAF}$ (the Image Reduction and Analysis Facility) routines were used to process the echelle spectroscopy data for flat and bias correction, wavelength calibration, and order merging. No flux calibration of the spectra was needed for our analysis.

\begin{table}
\caption{Summary of new spectroscopic observations by \emph{BFOSC}. The phases for these observations correspond to the ephemeris given in the \emph{Kepler} Eclipsing Binary Catalog.}
\begin{center}
\setlength{\tabcolsep}{2.0mm}{
\begin{tabular}{lccc}\hline\hline
UT-date	&	Exptime(s)	&	BJD	&	Phase	\\							\hline
2023-02-19	&	1800	&	2459995.369789 	&	0.04906	\\
2023-04-18	&	1800	&	2460053.255004 	&	0.19080 \\
2023-04-18	&	1800	&	2460053.275942 	&	0.19266	\\
2023-05-25	&	1800	&	2460090.183595 	&	0.47104	\\
2023-05-25	&	1800	&	2460090.204522 	&	0.47289	\\
\hline
\end{tabular}}
\end{center}
\label{tab:spec}
\end{table}

\section{Analysis and Results} \label{sec:result}

\subsection{Temperature of the primary star} \label{sec:Temperature}

The effective temperature, 7500 $\pm$ 60 K, was given by calculating the average temperature of the two components from \emph{LAMOST}. This is usually assigned to the primary star in the modeling procedure, $T_{eff1}$, which can be incorrectly estimated. Considering that the mixing of the binary components and the presence of $H_{\alpha}$ emission lines may affect the accuracy of the stellar atmospheric parameters, we try to determine the primary effective temperature by taking improved measures from the influence of the above two aspects. Based on the available spectroscopic data, V583 Lyr appears to be a single-line spectroscopic binary star. However, the lower resolution of the spectrum may be preventing us from distinguishing between the components of the system. Considering also that the mass ratio $q$ = $\frac{M_{2}}{M_{1}}$ and the temperature ratio $\frac{T_{2}}{T_{1}}$ of the primary and secondary stars are small, we consider the primary's contribution to be dominant. Therefore, we utilize the University of Lyon Spectroscopic Analysis Software (ULySS, \citealp{2009A&A...501.1269K}) to derive the atmospheric parameters by minimizing the $\chi^{2}$ value between the observed spectrum and a multidimensional grid of model spectra, which is constructed by interpolating the MILES library \citep{2006MNRAS.371..703S,2011A&A...531A.165P}. The synthetic spectra were produced using a grid of Kurucz model atmospheres \citep{1979ApJS...40....1K} under Local Thermodynamic Equilibrium (LTE) conditions. The models have been calculated by means of a statistical distribution function representation of the opacity of almost $10^{6}$ atomic lines.

This interpolator returns a spectrum for any temperature, metallicity, and gravity where each wavelength bin is calculated by interpolation over the entire reference library. It is constructed from three different sets of polynomials for the OBA, FGK, and M-type temperature ranges, and is linearly interpolated in overlapping regions. Each of these sets of polynomials is valid for a wide range of parameters, so it is a global interpolation. 

\begin{figure}[ht!]
\plotone{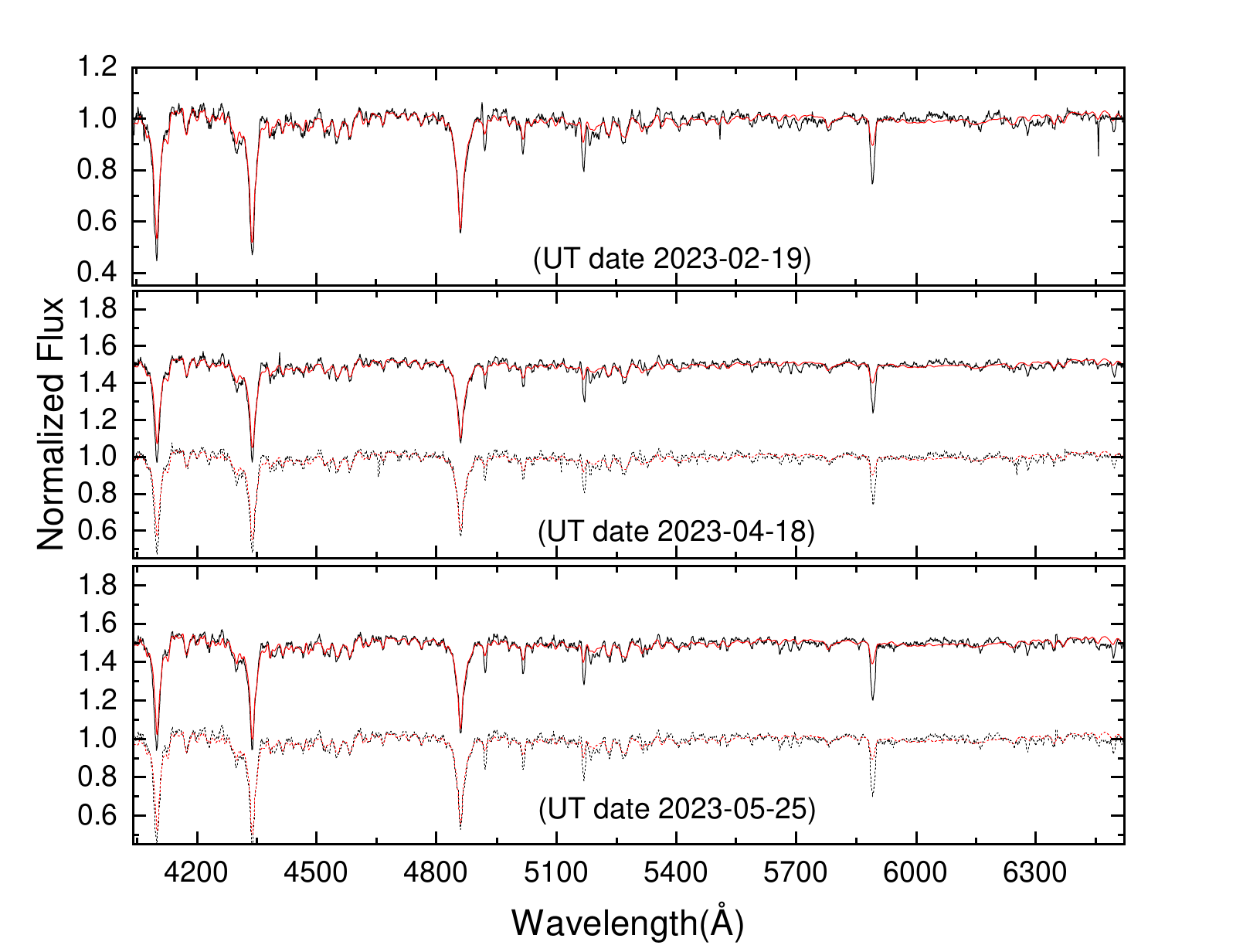}
\caption{Spectrum fitting for V583 Lyr. Refer to Table \ref{tab:spec-Teff} for the stellar parameters used for calculating the model spectrum.
\label{fig:UlySS-fit-results}}
\end{figure}

The shape of stellar spectra is mainly influenced by $T_{eff}$, $\log g$, and $[Fe/H]$. To determine the $T_{eff1}$ for V583 Lyr, the strong absorption lines that reliably trace $T_{eff1}$ were ultimately selected. The fit is performed using Equation \ref{ulyss fit} over the wavelength range 4000-6500Å, where the $H_{\alpha}$ emission lines are rejected.

\begin{equation*}\label{ulyss fit} 
    Obs(\lambda) = P_{n} \times G \bigotimes TGM(T_{eff}, \log g, [Fe/H], \lambda),
\tag{3.1} 
\end{equation*}
where Obs($\lambda$) is the observed one-dimensional spectral function of the wavelength ($\lambda$) ; $P_{n}$ is a multiplicative polynomial of degree $n$; $G$ is a Gaussian broadening function; and $TGM$ is a model spectrum as a function of $T_{eff}$, $\log g$, and $[Fe/H]$. The best-fitted results are shown in Figure \ref{fig:UlySS-fit-results}, and the atmospheric parameters are listed in Table \ref{tab:spec-Teff}. There are 343 values for V585 Lyr in the \emph{Gaia} DR3 BP/RP spectrum \citep{2022yCat.1355....0G} in a wide wavelength range from 336 to 1020 nm with 2 nm steps. The BP/RP spectrum of V583 Lyr estimated the effective temperature from the GSP-Phot Aeneas best library to be 9140.3 K, with lower (16\%) and upper (84\%) confidence limits of 8833.6 K and 9384.8 K, respectively. When dealing with A-type stars, it is important to consider the uncertainty in the effective temperature, which can be at least 100 K due to several factors, including continuum placement or spectrum normalization, uncertainties in microturbulence, and opacities used in the original model atmospheres. Therefore, based on the results listed in Table \ref{tab:spec-Teff} and the \emph{Gaia} DR3 BP/RP spectra, we have determined the mean temperature of $T_{eff1}$ to be 9000 K with an estimated uncertainty of about 350 K.

\begin{table}
\caption{Derived atmospheric parameters for the V583 Lyr.}
\begin{center}
\setlength{\tabcolsep}{2.0mm}{
\begin{tabular}{lcccc}\hline\hline
UT-date	&	Phase	&	$T_{eff}$	&	$\log g$	&	$[Fe/H]$	\\
 	&	 	&	 (K)	&	($cm s^{-2}$)	&	(dex)   \\
\hline
2023-02-19	&	0.049 	&	8970$\pm$180	&	4.20$\pm$0.1	&	0.35$\pm$0.06	\\
2023-04-18	&	0.191 	&	8980$\pm$200	&	4.04$\pm$0.1	&	0.34$\pm$0.06	\\
2023-04-18	&	0.193 	&	9000$\pm$200	&	4.06$\pm$0.1	&	0.33$\pm$0.07	\\
2023-05-25	&	0.471 	&	8970$\pm$160	&	4.24$\pm$0.1	&	0.39$\pm$0.05	\\
2023-05-25	&	0.473 	&	8970$\pm$160	&	4.22$\pm$0.1	&	0.37$\pm$0.05	\\
\hline
\end{tabular}}
\end{center}
\label{tab:spec-Teff}
\end{table}

\subsection{Photometric analysis} \label{sec:photo}

In this section, we analyzed the new photometric data from \emph{TESS} and reanalyzed the \emph{Kepler} data for V583 Lyr. The Wilson-Devinney (W-D; \citealp{1971ApJ...166..605W,1979ApJ...234.1054W,1994PASP..106..921W}) program was used with a detailed treatment of limb darkening, gravitational darkening, and reflection effects to obtain a comprehensive set of stellar and orbital parameters. Using the ephemeris $BJD_{0}$ = 55007.62275 (Barycentric Julian Date-2400000) given in the \emph{Kepler} Eclipsing Binary Catalog, we computed the phases and folded the light curve. The \emph{TESS} and \emph{Kepler} data were binned to 1000 and 600 points, respectively, for use in the binary modeling code.

\begin{figure}[ht!]
\plotone{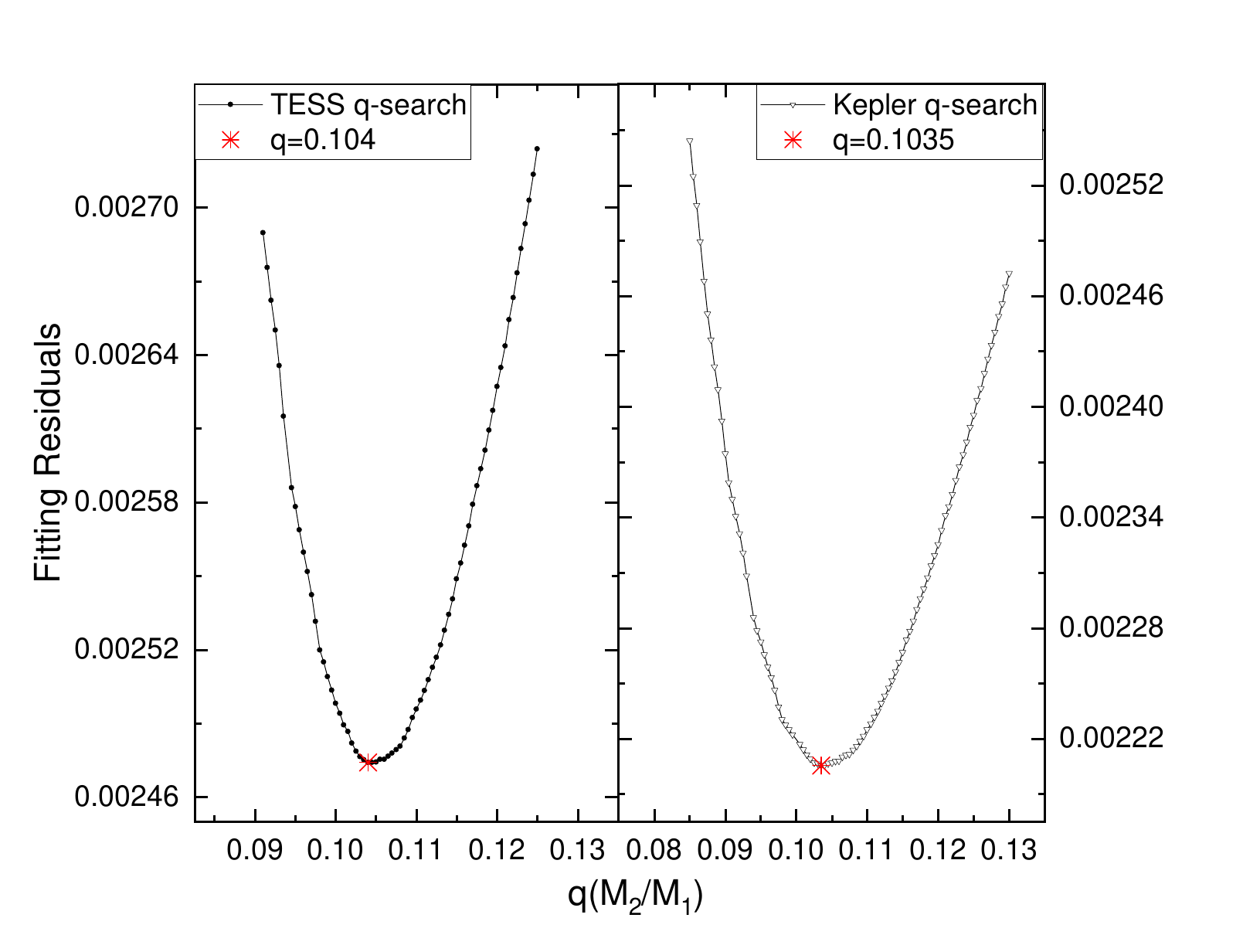}
\caption{The results of the least squares minimization to estimate the initial value of the mass ratio $q$. The fitting residuals are plotted for 0.07 $\leq$ $q$ $\leq$ 0.13 with the step of 0.0005. 
\label{fig:q-search}}
\end{figure}

Mode 5 (Semi-detached binary with the secondary star filled with its Roche Lobe) is chosen with a circular orbit (the orbital eccentricity $e=0$). The adjustable orbital parameters include orbital inclination ($i$), the surface temperature of the secondary component ($T_{eff2}$), the dimensionless potential of the primary component ($\Omega_{1}$), semi-major axis ($a$), phase shift ($\phi$), luminosity of the primary component ($L_{1}$), and third light ($l_{3}$). And pass-band limb-darkening coefficients were from \cite{1993AJ....106.2096V}'s table. 

\begin{figure}[ht!]
\begin{minipage}[t]{0.4\linewidth}
\centering
\includegraphics [width=3.8in]{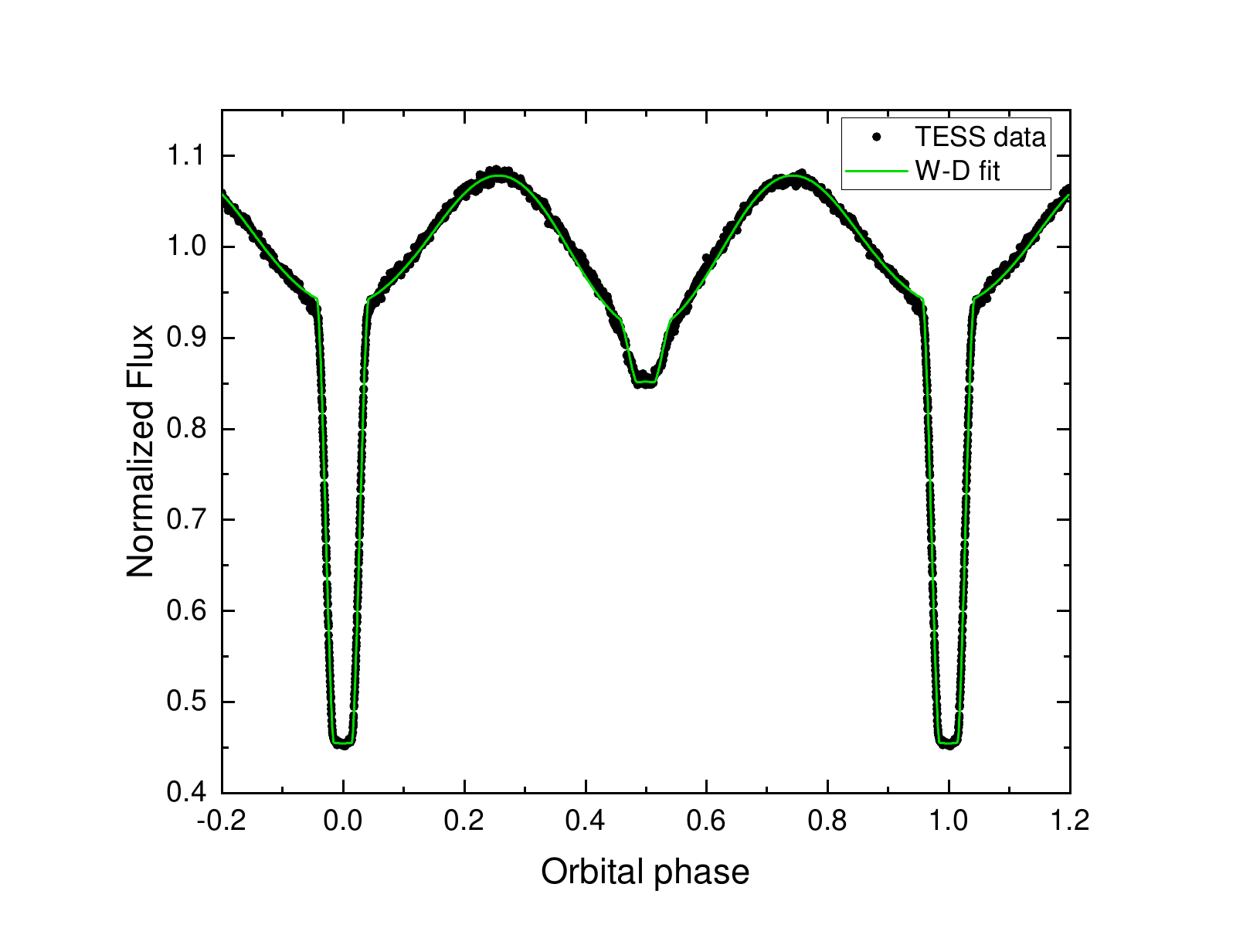}
\end{minipage}
\begin{minipage}[t]{0.4\linewidth}
\centering
\includegraphics [width=3.8in]{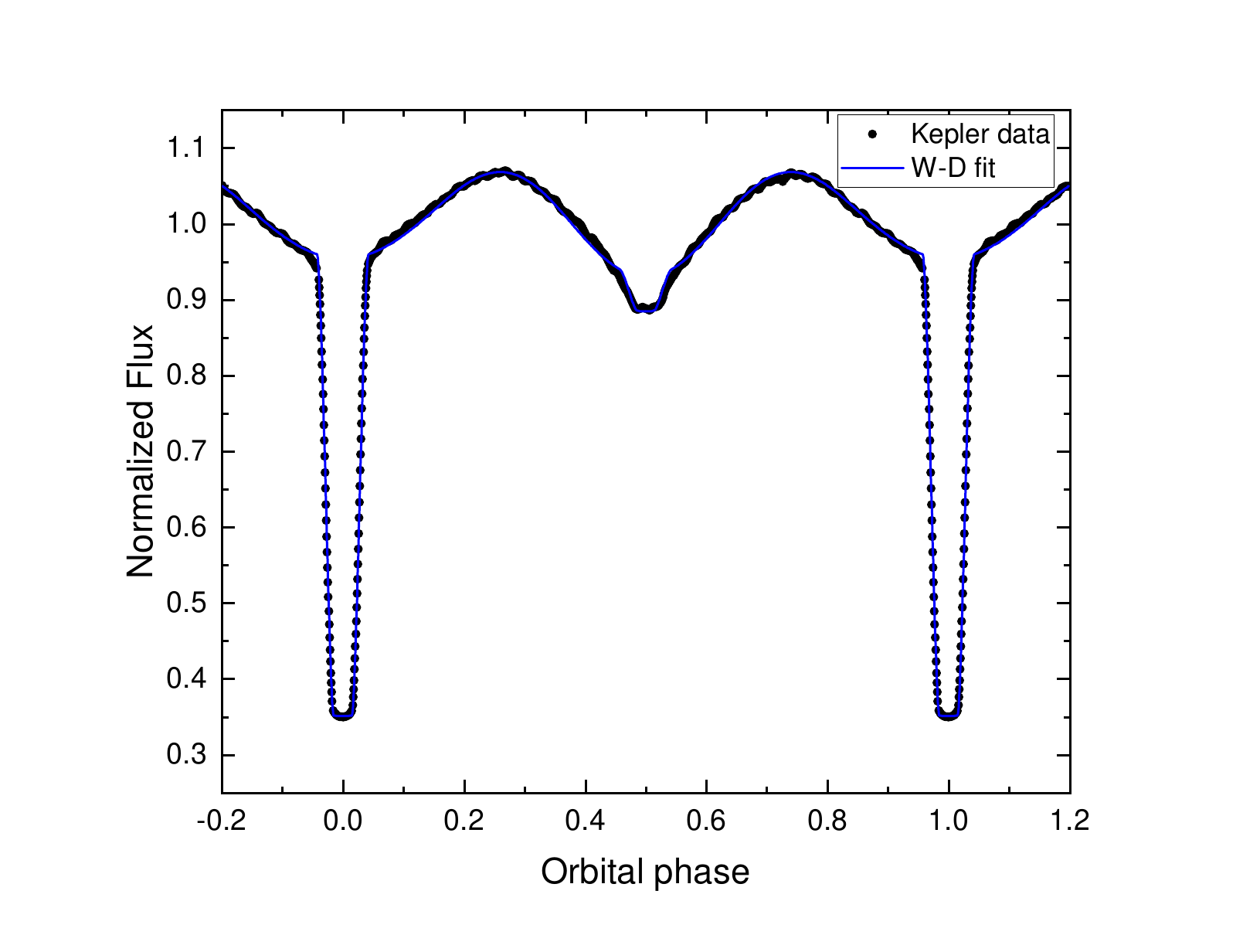}
\end{minipage}
\caption{Theoretical binary model fit of \emph{TESS} (left panel) and  \emph{Kepler} (right panel) in semi-detached mode.} 
\label{fig:wd-fit}
\end{figure}

\begin{figure}[ht!]
\plotone{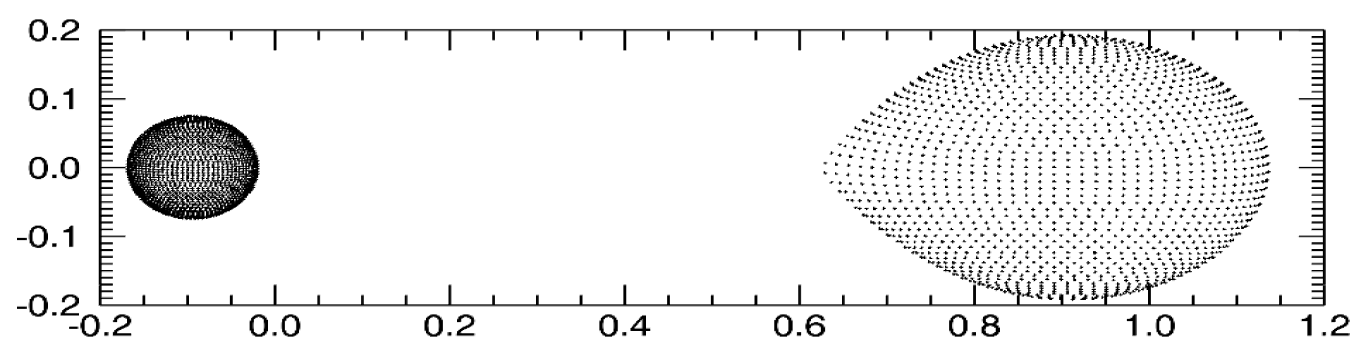}
\caption{The semi-detached configuration of V583 Lyr in the phase = 0.25. 
\label{fig:configuration-0.25}}
\end{figure}

The primary effective temperature is fixed as $T_{eff1}$=9000 $\pm$ 350 K obtained in Section \ref{sec:Temperature}. We determined the mass ratio ($q=M_{2}/M_{1}$) by using the q-search \footnote{q-search is performed with the W-D program to search for the best fit mass ratio of the binary model in a given range of $q$ values.} method with the step of 0.0005 in the range 0.07 $\leq$ $q$ $\leq$ 0.13. The results of the q-search are plotted in Figure \ref{fig:q-search} for \emph{TESS} (left panel) and \emph{Kepler} (right panel) photometry, which obtained the close search results of $q$ = 0.1035 $\pm$ 0.0022 and $q$ = 0.1040 $\pm$ 0.0016, and also close to the result of $q$ = 0.0962 $\pm$ 0.0024 obtained by \cite{Zhang_2020}. Based on the above results, we finally determine the best estimate for the mass ratio of V583 Lyr to be $q$ = 0.1 $\pm$ 0.004.

We then utilized both \emph{TESS} and \emph{Kepler} data for binary model fitting, yielding the orbital parameters and relative physical parameters listed in Table \ref{tab:wd-results}. The temperature ratio $T_{2}/T_{1}$, relative radii $r/a$, orbital inclination $i$, and mean density $\rho_{1,2}$ obtained in our results generally align with those of \cite{Zhang_2020}, who solely analyzed \emph{Kepler} data. The bolometric gravity darkening coefficient of the primary star ($g_{1}$) was set to 1.0 for radiative atmospheres from \cite{1924MNRAS..84..665V}, and the bolometric albedo value ($A_{1}$) was set to 1.0, from \cite{1969AcA....19..245R}. While $g_{2}$ and $A_{2}$ of the secondary star are adjusted by the binary model because the detailed reflection effect and the Roche geometry of the evolved secondary star can be taken into account. We obtained the results $g_{2}$ = 0.313 $\pm$ 0.001 and $A_{2}$ = 0.715 $\pm$ 0.006 from \emph{TESS} data and $g_{2}$ = 0.298 $\pm$ 0.003 and $A_{2}$ = 0.775 $\pm$ 0.008 from \emph{Kepler} data. The final photometric solutions from \emph{Kepler} and \emph{TESS} photometry are listed in Table \ref{tab:wd-results}. 

The luminosity ratio of $L_{1,2}/(L_{1} + L_{2})$ varies between the \emph{Kepler} and \emph{TESS} bands. In the \emph{Kepler} band, $L_{1}/(L_{1} + L_{2})$ comprises approximately 58\% while $L_{2}/(L_{1} + L_{2})$ comprises roughly 42\%. Conversely, in the \emph{TESS} band, $L_{1}/(L_{1} + L_{2})$ amounts to 46\%, and $L_{2}/(L_{1} + L_{2})$ amounts to 54\%. This illustrates the secondary star's superluminosity, potentially resulting from variances in photometric band ranges between \emph{Kepler} and \emph{TESS}. Finally, the light curves that resulted from the binary model's fitting in both \emph{Kepler} and \emph{TESS} bands are displayed in Figures \ref{fig:wd-fit}, and the semi-detached configuration is presented in Figure \ref{fig:configuration-0.25}. Our results of the photometric analysis give the mean relative radius of the gainer as $R_{1}/a$ $\approx$ 0.074. It is below the $\omega_{min}$ curve based on the semi-analytical ballistic calculations of \cite{1975ApJ...198..383L}, indicating the formation of a permanent disk (e.g., TT Hya, \citealp{1993MNRAS.262..220V,2007ApJ...656.1075M}).

\begin{table}
\caption{The orbital and physical parameters resulting from the best-fitting binary model to \emph{Kepler} and \emph{TESS} photometry of V583 Lyr. The numbers in parentheses are the errors on the last bits of the data. $f_{1,2}$ represents the filling factors, which is the ratio of the stars' volume to their Roche lobe volume ($V_{star}$/$V_{RL}$).}
\begin{center}
\setlength{\tabcolsep}{2.0mm}{
\begin{tabular}{lcc}\hline\hline
Parameters & \emph{TESS} & \emph{Kepler} \\
\hline
$P_{orb} (d)$	&	11.2580	&	11.2580	\\
mode	&	 semi-detached	&	 semi-detached	\\
$i(deg)$	&	85.456(21)	&	85.509(55)	\\
$q=M_{2}/M_{1}$	&	0.104	&	0.1035	\\
$T_{2}/T_{1}$	&	0.5416(9)	&	0.5316(18)	\\
$L_{1}/(L_{1}+L_{2})$	&	0.4629(4)	&	0.5804(9)	\\
$L_{2}/(L_{1}+L_{2})$	&	0.5371(4)	&	0.4196(9)	\\
$g_{1}$	&	1.0(assumed)	&	1.0(assumed)	\\
$g_{2}$	&	0.313(1)	&	0.298(3)	\\
$A_{1}$	&	1.0(assumed)	&	1.0(assumed)	\\
$A_{2}$	&	0.715(6)	&	0.775(8)	\\
$\Omega_{1}$	&	13.55(5)	&	13.82(14)	\\
$\Omega_{2}$	&	1.971(11)	&	1.970(6)	\\
$r_{pole1}/a$	&	0.0743(3)	&	0.0730(8)	\\
$r_{pole2}/a$	&	0.1921(12)	&	0.1918(6)	\\
$r_{point1}/a$	&	0.0743(3)	&	0.0730(8)	\\
$r_{point2}/a$	&	0.2856(12)	&	0.2852(6)	\\
$r_{side1}/a$	&	0.0743(3)	&	0.0730(8)	\\
$r_{side2}/a$	&	0.1997(12)	&	0.1994(6)	\\
$r_{back1}/a$	&	0.0743(3)	&	0.0730(8)	\\
$r_{back2}/a$	&	0.2307(12)	&	0.2305(6)	\\
$R_{2}/R_{1}$	&	2.794(6)	&	2.707(21)	\\
$f_{1}$	&	2.163(13)	&	2.033(34)	\\
$f_{2}$	&	100.00(6)	&	100.00(5)	\\
$\rho_{1}(\rho_{\odot})$	&	0.2331(9)	&	0.2476(27)	\\
$\rho_{2}(\rho_{\odot})$	&	0.0011(2)	&	0.0011(1)	\\
\hline
\end{tabular}}
\end{center}
\label{tab:wd-results}
\end{table}

\subsection{Radial Velocities and Mass function} \label{sec:RV}

In this section, we attempt to measure radial velocities (RVs) to determine the fundamental properties of V583 Lyr. The heliocentric correction is done for \emph{BFOSC} spectra using the $\mathtt{IRAF}$ $\mathtt{rvcorrect}$ task, and for \emph{LAMOST} LRS spectra the vacuum wavelength has been converted to air wavelength using the refractive index of \cite{Ciddor:96}. The spectral lines of the secondary star are much sharper in our optical spectra than those of the primary star. This is because the lines of the primary are highly broadened by the star's rotation and distorted by the presence of the accretion disk. Therefore, the lines of the secondary star are more accurate tracers of the orbital motion, resulting in more precise orbital elements compared to those derived from the lines of the primary. 

\begin{table}
\caption{Heliocentric radial velocities of the gainer obtained from the V583 Lyr spectra of \emph{LAMOST} (2017) and \emph{BFOSC} (2023), and their respective errors, using the Gaussian function. Residuals from the RV-fitting function are indicated as O-C.}
\begin{center}
\setlength{\tabcolsep}{1.6mm}{
\begin{tabular}{lcccc}\hline\hline
Phase	&	Velocity	&	error	&	O-C  &  Template spectra	\\
 	&	 (km $s^{-1}$)	&	(km $s^{-1}$)	&	(km $s^{-1}$)  &   	\\
\hline 
0.049	&	-30.493 	&	15.332 	&	15.370 	&	19951104-HD50692	\\
0.049	&	-34.777 	&	15.361 	&	11.086 	&	19971116-HD50692	\\
0.049	&	-27.737 	&	15.253 	&	18.126 	&	20011016-HD50692	\\
	&		&		&		&		\\
0.191	&	48.408 	&	19.188 	&	-14.309 	&	19951104-HD50692	\\
0.191	&	45.327 	&	19.100 	&	-17.390 	&	19971116-HD50692	\\
0.191	&	52.896 	&	18.920 	&	-9.821 	&	20011016-HD50692	\\
	&		&		&		&		\\
0.193	&	60.376 	&	18.964 	&	-4.039 	&	19951104-HD50692	\\
0.193	&	56.948 	&	18.879 	&	-7.467 	&	19971116-HD50692	\\
0.193	&	63.216 	&	18.967 	&	-1.199 	&	20011016-HD50692	\\
	&		&		&		&		\\
0.471	&	191.938 	&	21.432 	&	7.111 	&	19951104-HD50692	\\
0.471	&	188.023 	&	21.557 	&	3.196 	&	19971116-HD50692	\\
0.471	&	194.004 	&	21.543 	&	9.177 	&	20011016-HD50692	\\
	&		&		&		&		\\
0.473	&	205.553 	&	21.288 	&	21.166 	&	19951104-HD50692	\\
0.473	&	202.264 	&	21.294 	&	17.877 	&	19971116-HD50692	\\
0.473	&	208.916 	&	21.461 	&	24.529 	&	20011016-HD50692	\\
	&		&		&		&		\\
0.663	&	57.669 	&	10.241 	&	-10.864 	&	19951104-HD50692	\\
0.663	&	54.141 	&	10.362 	&	-14.392 	&	19971116-HD50692	\\
0.663	&	61.271 	&	10.308 	&	-7.262 	&	20011016-HD50692	\\
\hline
\end{tabular}}
\end{center}
\label{tab:rv1}
\end{table}

\begin{figure}[ht!]
\plotone{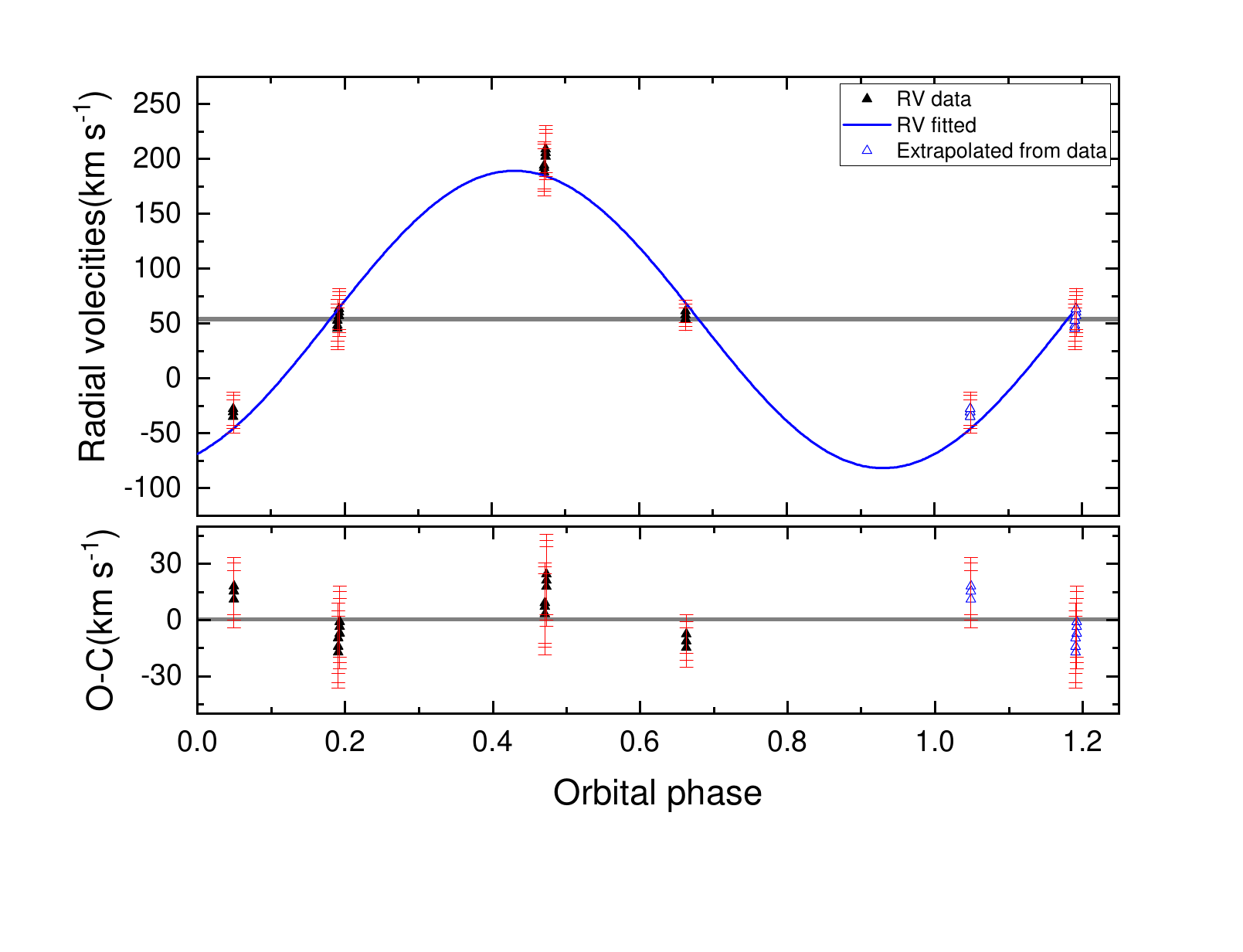}
\caption{Upper panel: Velocity curve of V583 Lyr based on elements from Table \ref{tab:rv1} and the best-fitting solution. The horizontal dashed line marks the corresponding systemic velocity. Bottom panel: residuals from the fit. 
\label{fig:RV-FIT}}
\end{figure}

The radial velocities were determined using the cross-correlation technique with an $\mathtt{IRAF}$ radial velocity package $\mathtt{xcsao}$\citep{1979AJ.....84.1511T,1992ASPC...25..432K,1998ASPC..145...93M}. We selected a group of template spectra with known velocities from ELODIE library \footnote{\url{http://atlas.obs-hp.fr/elodie/}} \citep{2001A&A...369.1048P}, including three spectra from HD50692 (G0V, $v\sin{i}$ = 15 km $s^{-1}$). The observed spectra were cross-correlated with the template spectrum to measure the radial velocities of the secondary star. The region near $H_{\alpha}$ was omitted during this process. The observations and the derived heliocentric radial velocities against the template are listed in Table \ref{tab:rv1}. 
In Figure \ref{fig:RV-FIT}, the $RVs$ are sorted by phase and fitted with a sine function through a Marquart-Levenberg method \citep{doi:10.1137/0111030}: 

\begin{equation*}\label{RV-fit}
    RV = \gamma +K\sin{[2\pi(\phi+\phi_{0})]}
\tag{3.3} 
\end{equation*}
A system velocity $\gamma$ = 53.7 $\pm$ 3 km $s^{-1}$, the semi-amplitude of the secondary $K_{2}$ = 135.5 $\pm$ 6 km $s^{-1}$ and phase shift $\phi_{0}$ = -0.68 $\pm$ 0.01 are estimated. The residuals from the RV-fitting function are indicated as O-C, listed in Table \ref{tab:rv1}.

\begin{table}
\caption{Absolute parameters of V583 Lyr compared to the results of \cite{Zhang_2020}.}
\begin{center}
\setlength{\tabcolsep}{1.6mm}{
\begin{tabular}{lcc}\hline\hline
Parameters	&	This work	&	 \cite{Zhang_2020}	\\
\hline
$T_{eff1}$(K)	&	9000 $\pm$ 350	&	7484 $\pm$ 58	\\
$T_{eff2}$(K)	&	4780 $\pm$ 100	&	4366 $\pm$ 57	\\
$M_{1}/M_{\odot}$	&	3.56 $\pm$ 0.5	&	1.58 $\pm$ 0.49	\\
$M_{2}/M_{\odot}$	&	0.36 $\pm$ 0.02	&	0.153 $\pm$ 0.051	\\
$R_{1}/R_{\odot}$	&	2.4 $\pm$ 0.2	&	1.90 $\pm$ 0.28	\\
$R_{2}/R_{\odot}$	&	6.9 $\pm$ 0.4	&	5.14 $\pm$ 0.79	\\
\hline
\end{tabular}}
\end{center}
\label{tab:absolute paraters}
\end{table}

We calculated the mass function $f(m)$ defined for a single-lined spectroscopic binary as:
\begin{equation*}\label{f(m)}
    f(m)=\frac{M_{1}\sin^{3}{i}}{(1+q)^{2}}= \frac{P_{orb}}{2\pi G}K_{2}^{3},
\tag{3.4}
\end{equation*}
where $P_{orb}$ = 11.2580 $d$ is the orbital period, the orbital inclination $i$ and the mass ratio $q_{phot}$ are the photometric solutions from Section \ref{sec:photo}.
The mass of the primary star is calculated from Equation \ref{f(m)} as $M_{1}$ = 3.56 $\pm$ 0.5 $M_{\odot}$, and $M_{2}$ = $M_{1}\times q$ = 0.36 $\pm$ 0.02 $M_{\odot}$ is derived from the mass ratio $q$ $\approx$ 0.1.
Then the semi-major axis is calculated using the third Kepler's low:

\begin{equation*}\label{third Kepler's low}
    \frac {a^3} {P^2} =\frac {G} {4\pi^2} (M_1+M_2) 
\tag{3.5}
\end{equation*}
to be $a$ = 33.34 $\pm$ 0.18$R_{\odot}$. The estimated absolute parameters are listed in Table \ref{tab:absolute paraters} in comparison with the results of \cite{Zhang_2020}. They estimated the absolute physical characteristics of stars using the $\rho-T$ method (density-temperature method introduced in \citealt{2017MNRAS.466.1118Z}) in their study.

\subsection{ \texorpdfstring {$H_{\alpha}$}{} profiles and accretion disk} \label{sec:H profiles}

\begin{figure}[ht!]
\plotone{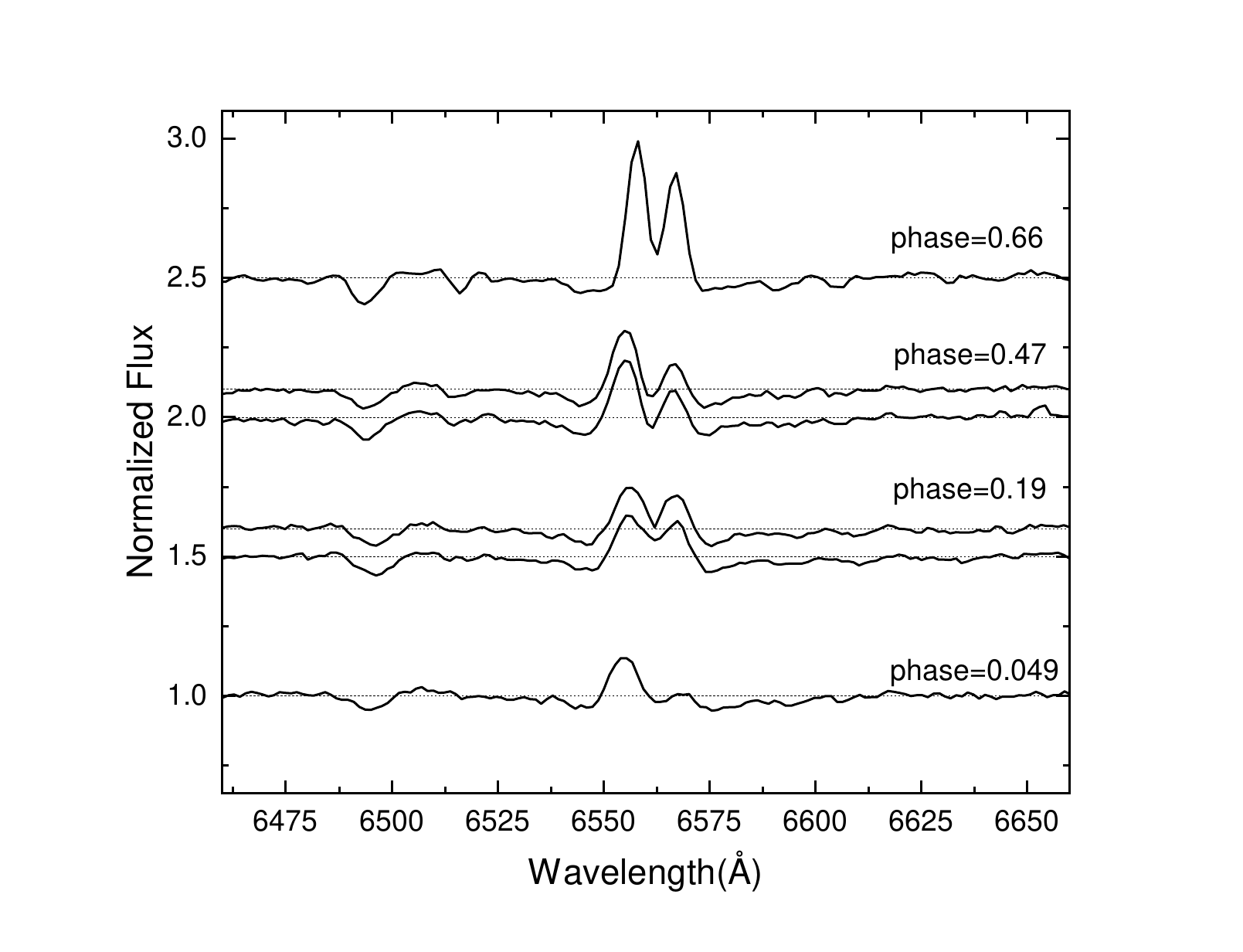}
\caption{All the spectra show $H_{\alpha}$ profile. Fluxes are normalized to the continuum and heliocentric corrections have been applied. 
\label{fig:Ha-spectrum}}
\end{figure}

\begin{figure}[ht!]
\plotone{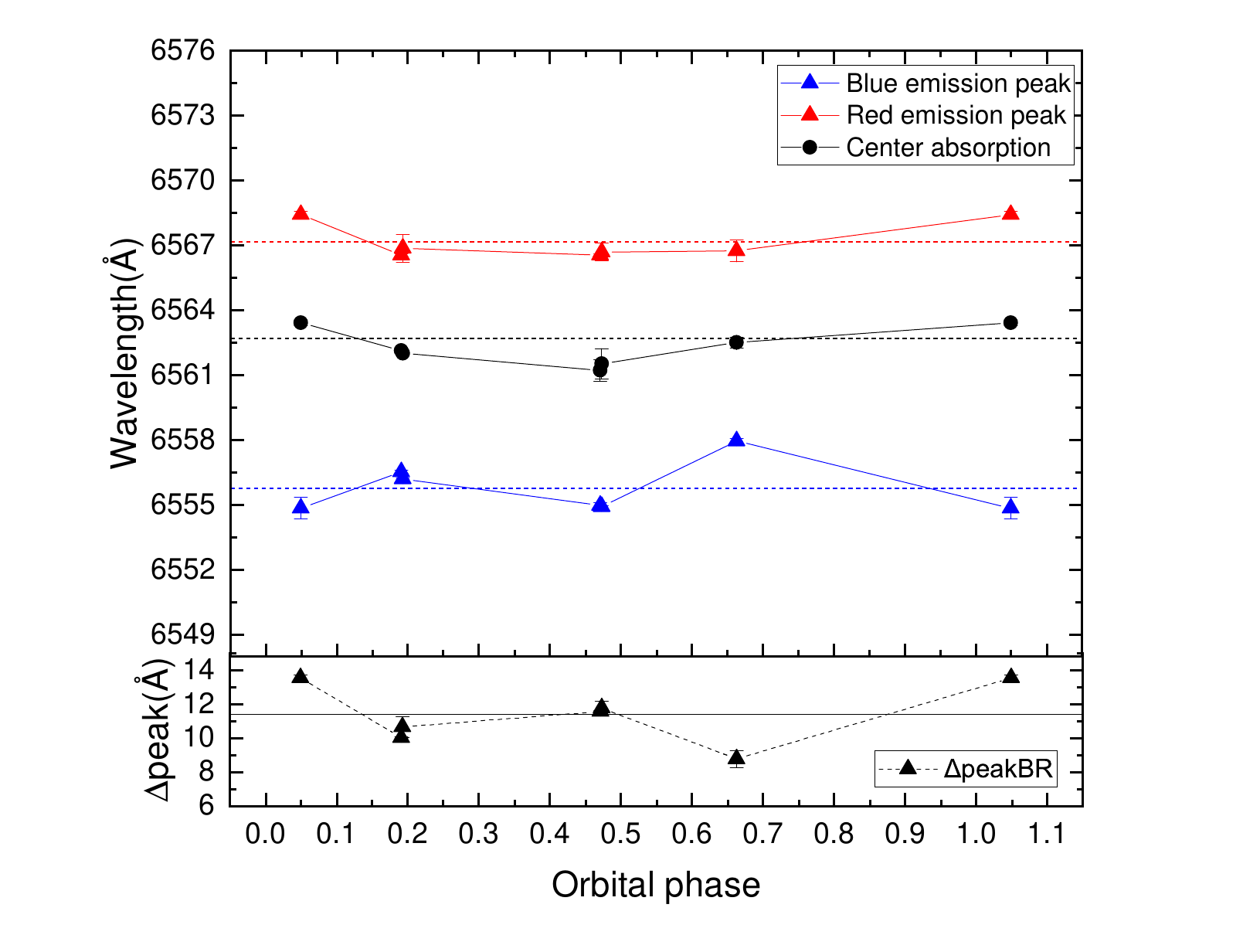}
\caption{The shifts of the double peaks and the central absorption are plotted against the variation of the average peak separation. 
\label{fig:Ha-profiles-variation}}
\end{figure}

\begin{figure}[ht!]
\plotone{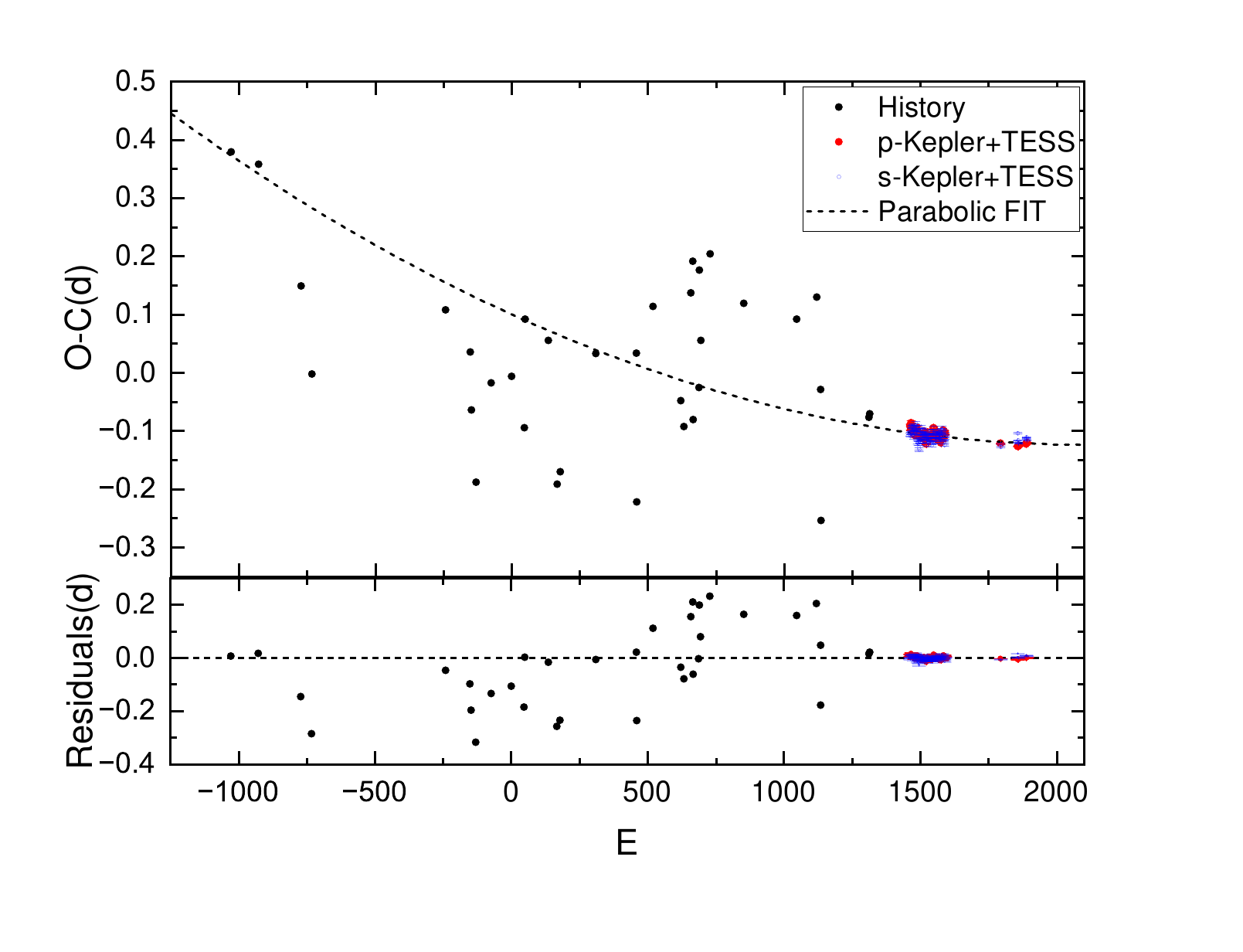}
\caption{The theoretical fit to the O-C data for V583 Lyr. The top panel shows the parabolic fit (dashed line), and the bottom panel shows the residuals from the fit. 
\label{fig:O-C}}
\end{figure}

The double-peaked emission is the most intriguing aspect of the $H_{\alpha}$ profile, which is mainly due to Thomson scattering of the light from the primary star. The accretion disk scatters light from the primary star, creating a wide emission line. This is superimposed on the absorption profile in the stellar atmosphere of the primary, resulting in the double peak emission profile. The broad wings of the emission profile extend from 6547 to 6574 Å. The variability of the $H_{\alpha}$ line profile with orbital phase is illustrated in Figure \ref{fig:Ha-spectrum}. The properties of $H_{\alpha}$ emission profile are listed in Table \ref{tab:Ha}. From the currently available spectroscopic observations plotted in Figure \ref{fig:Ha-spectrum}, we can see the $H_{\alpha}$ blue-peak emission is larger than the red peak. The ratio $B/R$ \footnote{$B/R$, historically known as $V/R$, is still widely used.} of the $H_{\alpha}$ profile is calculated and listed in Table \ref{tab:Ha}. 

The variation of the center wavelength of the blue and red peaks with the orbital period is shown in Figure \ref{fig:Ha-profiles-variation}. The red peak shifts along with the central absorption line, and the average peak separation is 11.067 $\pm$ 0.297 Å. This value was used to determine the velocity of a particle located on the accretion disk rim, which is $V_{rim}$ = 253 $\pm$ 10 $km s^{-1}$. The spectral data in Figure \ref{fig:Ha-spectrum} and Table \ref{tab:Ha} show that the blue emission during the primary eclipse (phase = 0.049) is unexpectedly strong, while the red emission is comparatively weaker, as indicated by the double peaked emission intensities. The excess absorption in the core of $H_{\alpha}$ during these phases could be attributed to the eclipse effect, which suggests the possible presence of cooler circumstellar matter surrounding the secondary star. Figure \ref{fig:Ha-profiles-variation} shows the shifts of the blue and red peaks, as well as the central absorption, in comparison to the variation of the average peak separation $\Delta peak$. The red peak changes consistently with absorption, but the change in the blue peak is significantly different at phase 0.049. Variations in the emission lines of Algol binaries typically occur around phase 0.0 when the secondary star eclipses the disk rotating with the primary star. This is most obvious in totally eclipsing systems. The $H_{\alpha}$ profile for phase 0.049 (post eclipse) in Figure \ref{fig:Ha-spectrum} resembles a disk system during its egress phase interval, similar to TT Hya \citep{2007ApJ...656.1075M} or AU Mon \citep{2012ApJ...760..134A}. At phase 0.66, the gas stream approaches the observer. If its material emits in $H_{\alpha}$, it could contribute to the disk emission. Phases around 0.2 show comparable intensity in the blue-red emission peaks, while the blue emissions are stronger following the secondary eclipse. During phase 0.5-0.6, the blue emission is stronger than the red emission. Additionally, the central absorption deepens when leaving the secondary eclipse. Studies \citep{2007ApJ...656.1075M,2012ApJ...760..134A} on accretion disks suggest that this is due to the extra absorption of the secondary's light in the reappearing part of the disk and the eclipse of the secondary star. 

The direction of mass transfer is discovered to be consistent with the period change analyzed by the O-C method. We have collected all eclipse times for V583 Lyr, consisting of the new data from \emph{Kepler} and \emph{TESS} photometry, which are listed in Table \ref{tab:min} in the Appendix. The upward O-C curve in Figure \ref{fig:O-C} indicates that V583 Lyr is currently experiencing an increase in the orbital period due to mass transfer from a low-mass donor to a more massive gainer: 

\begin{equation*}\label{o-c}
    O-C=5.064 \times 10^{-8} (\pm 0.5 \times 10^{-8} )\times E^{2} - 2.1 \times 10^{-5} (\pm 1.1 \times 10^{-5} )\times E + 0.100(\pm 0.007 ).
\tag{4.1}
\end{equation*}
where $E$ represents the epochs of the primary minima in cycles. The rate of mass transfer is estimated by assuming conservative mass transfer \footnote{The mass transfer may not be conservative due to the system's uncertain location in the r-q diagram. However, for estimation purposes, a conservative approach can be assumed since the accurate determination of the systemic mass loss rate is currently uncertain. Where the r-q diagram for interacting binaries shows the fractional radius of the mass gainer (R/a) plotted against the mass ratio ($q=M_{2}/M_{1}$). According to \cite{2022AAS...24020508P}, the r-q diagram can help determine whether the gas stream has a direct or tangential impact, or misses the mass gainer to form an accretion disk.}, and the rate of period change is calculated from $\frac{\dot{P}}{P}$:

\begin{equation*}\label{period change }
    \frac{\dot{P}}{P}  = 5.064\times10^{-8}\times2\times365.24219/(11.2580)^{2} = 2.9\times10^{-7} d\cdot yr^{-1}, 
\tag{4.2}
\end{equation*}
and for the conservative mass transfer:
\begin{equation*}\label{mass transfer}
    \frac{\dot{P}}{P}=\frac{3\dot{M_{1}}(M_{1}-M_{2})}{M_{1}M_{2}} 
\tag{4.3}
\end{equation*}
So the matter accreted by the primary star is calculated to be a rate at $\dot{M_{1}}$ = 3.384 $\times10^{-8}$ $M_{\odot}\cdot yr^{-1}$.

\begin{table}
\caption{The properties of $H_{\alpha}$ emission profile. The center wavelength $B(center)$, $R(center)$, and $C(center)$ $H_{\alpha}$ emission profile, the intensities of the blue and red emission peaks normalized to the underlying continuum $I_{B}$/$I_{cont}$ and $I_{R}$/$I_{cont}$, the intensity of the central absorption $I_{C}$/$I_{cont}$, peak separation $\Delta peak_{BR}$, and intensity ratio of the emission peaks $B/R$.}
\begin{center}
\setlength{\tabcolsep}{1.6mm}{
\begin{tabular}{lccccccccc}\hline\hline
Phase	&	Instrument	&	$B(center)$	&	$I_{B}$/$I_{cont}$	&	$R(center)$	&	$I_{R}$/$I_{cont}$	&	$C(center)$	&	$I_{C}$/$I_{cont}$	&	$\Delta peak_{BR}$	&	$B/R$	\\
	&		&	(\AA)	&	 	&	(\AA)	&	 	&	(\AA)	&	 	&	(\AA)	&	 	\\
\hline
0.04906	&	BFOSC	&	6554.85$\pm$0.5	&	1.13	&	6568.42$\pm$0.15	&	1.01	&	6563.42$\pm$0.12	&	0.99	&	13.57$\pm$0.15	&	1.13	\\
0.19080	&	BFOSC	&	6556.53$\pm$0.07	&	1.14	&	6566.56$\pm$0.05	&	1.13	&	6562.14$\pm$0.08	&	1.06	&	10.03$\pm$0.05	&	1.02	\\
0.19266	&	BFOSC	&	6556.21$\pm$0.04	&	1.15	&	6566.87$\pm$0.63	&	1.12	&	6562.01$\pm$0.03	&	1.01	&	10.66$\pm$0.63	&	1.03	\\
0.47104	&	BFOSC	&	6554.98$\pm$0.13	&	1.20	&	6566.56$\pm$0.05	&	1.09	&	6561.23$\pm$0.50	&	0.96	&	11.58$\pm$0.05	&	1.10	\\
0.47289	&	BFOSC	&	6554.92$\pm$0.05	&	1.21	&	6566.7$\pm$0.4	&	1.09	&	6561.53$\pm$0.70	&	0.98	&	11.78$\pm$0.4	&	1.11	\\
0.66266	&	LAMOST	&	6557.97$\pm$0.11	&	1.48	&	6566.75$\pm$0.5	&	1.36	&	6562.51$\pm$0.26	&	1.09	&	8.78$\pm$0.5	&	1.09	\\
\hline
\end{tabular}}
\end{center}
\label{tab:Ha}
\end{table}

\subsection{Pulsation analysis} \label{sec:Pulsation}

\begin{figure}[ht!]
\plotone{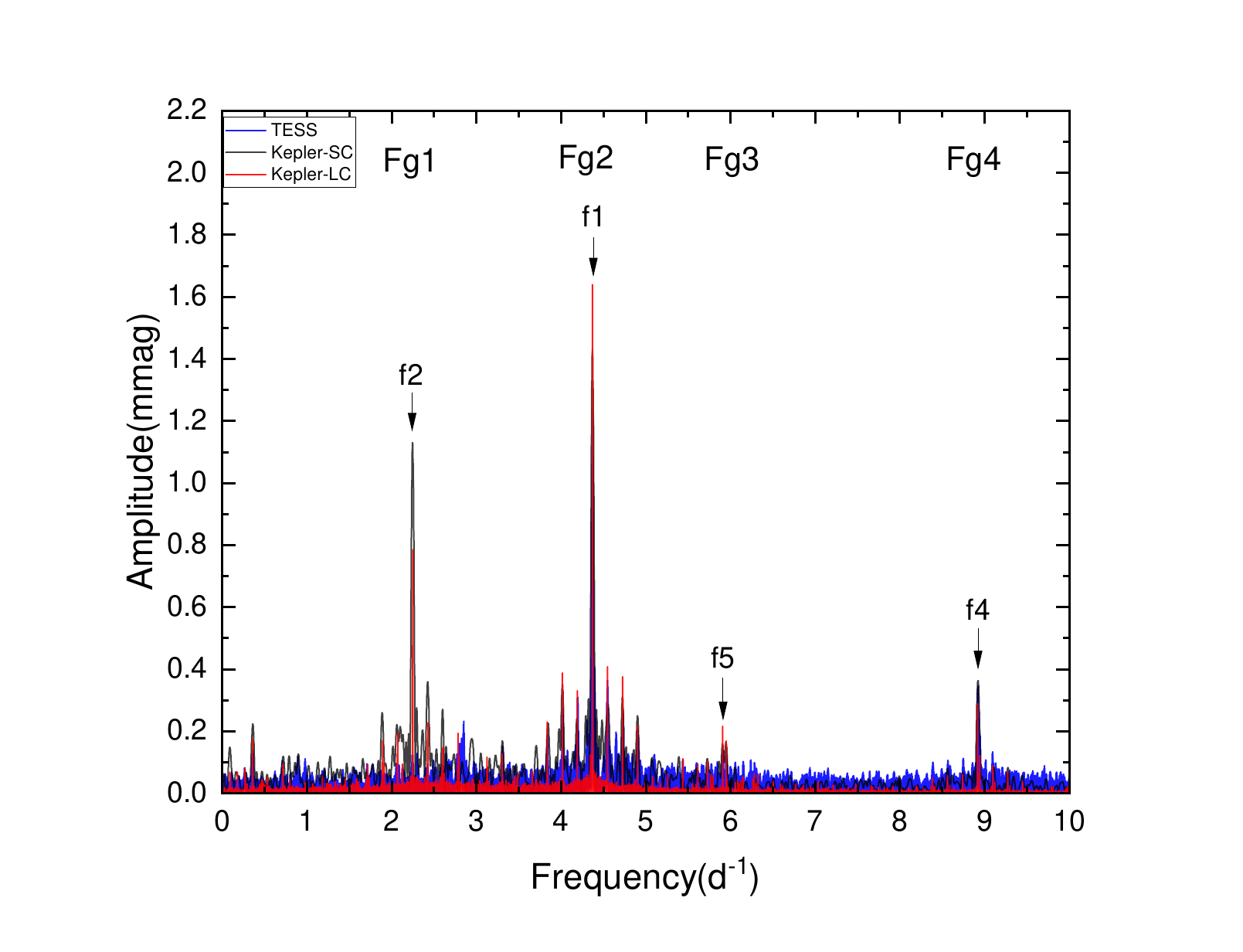}
\caption{Comparison of the Fourier periodograms of the \emph{TESS} (blue) and \emph{Kepler} -SC (black), -LC (red) light curves of V583 Lyr. There are no pulsations at higher frequencies up to the Nyquist frequency. The frequency groups are labeled as four groups with Fg1, Fg2, etc.  
\label{fig:Frequencies-coadd}}
\end{figure}

The frequency analysis of V583 Lyr was conducted using the long cadence (LC) data from \emph{Kepler} quarters 0 to 17 with Period04 \citep{2004IAUS..224..786L}. The LC observations were obtained over a continuous run of about 4 years (1,470 days) and covered nearly 130 full orbital cycles. The g-mode pulsation were initially deduced based on the range of frequencies and the mass of the primary star estimated in Section \ref{sec:RV}. The features of combination and group frequencies can be used to identify $g$ mode pulsations, which can have so small frequency separations that data sets of less than 1 year may not be able to distinguish individual pulsation frequencies. Examples of this can be found in the $\gamma$ Dor-$\delta$ Sct stars KIC 11145123 \citep{2014MNRAS.444..102K,2015A&A...574A..17V} and KIC 9244992 \citep{2014MNRAS.444..102K}, as well as in the catalog of $\gamma$ Dor stars presented by \cite{2020MNRAS.491.3586L}. Therefore, the \emph{Kepler} LC data set, which is more credible than the short cadence (SC) data for understanding the $g$ mode pulsation, was chosen in this work.

To disentangle the eclipse and pulsation in V583 Lyr, we use the methods from our early work on RZ Hor \citep{2023PASJ...75..732Z}. The frequency resolution was calculated as $\delta f_{LC}$ = $\frac{1.5}{\Delta T}$ $\approx$ 0.001 $d^{-1}$, and the Nyquist frequency is $f_{Ny}$ = 24.464 $d^{-1}$, where $\Delta T$ represents the length of the observation time. The signal-to-noise ratio (SNR) threshold for frequencies given against \emph{Kepler} photometric data is $S/N_{Kepler}$ = 5.4 \citep{2015MNRAS.448L..16B}. To compare the periodogram of the \emph{Kepler} (SC and LC data) and \emph{TESS} observations, we plot them together in Figure \ref{fig:Frequencies-coadd}. The three data sets show close agreement for the four major frequencies marked with arrows. The pulsation frequencies are concentrated in the low-frequency region ($f$ \textless 9 $d^{-1}$). 

\begin{figure}[ht!]
\plotone{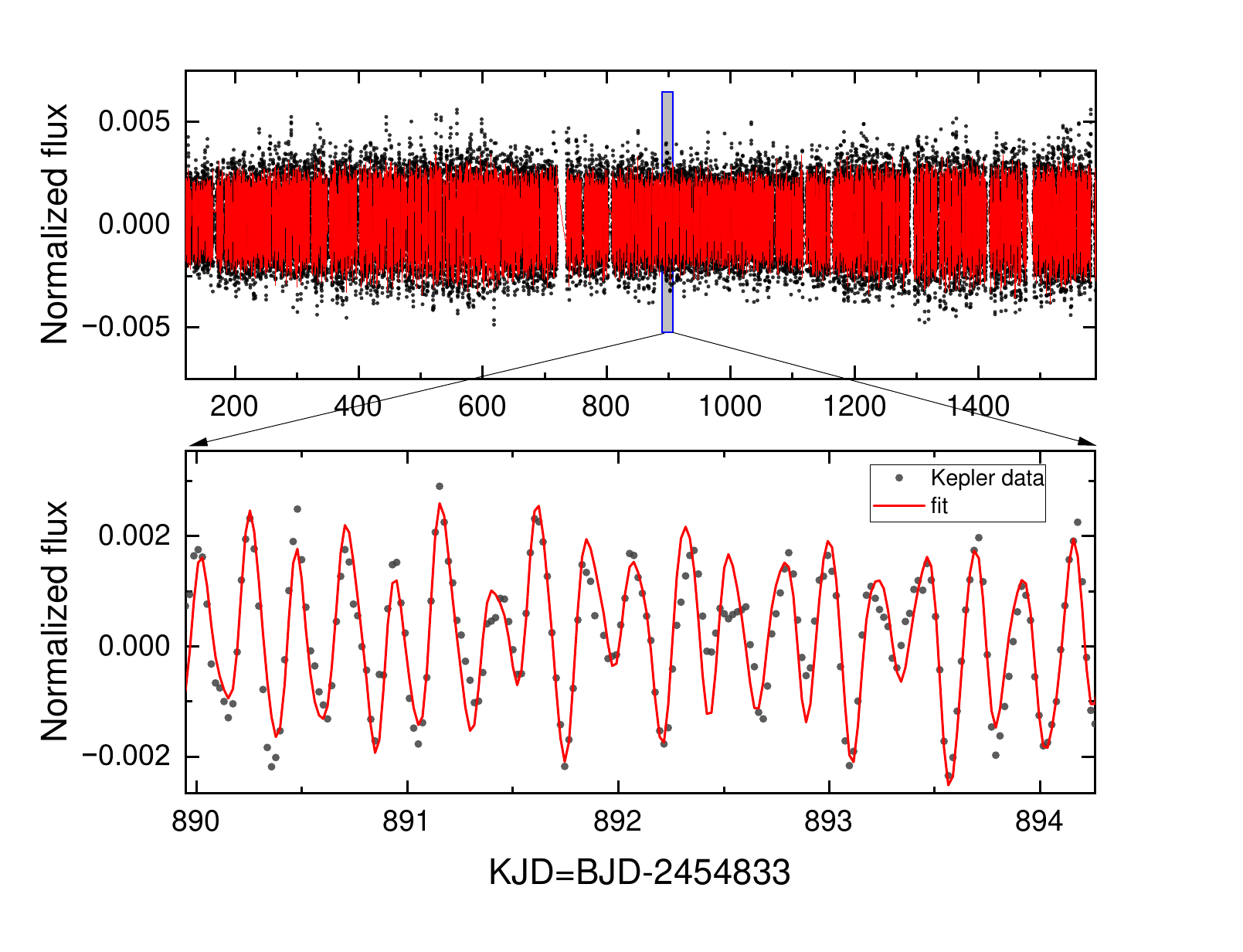}
\caption{Multi-frequency fit for V583 Lyr. The residuals after subtracting the eclipses in the \emph{Kepler} (LC) observations and the synthetic curve. The lower panel presents a portion of the residuals denoted with the inset box in the upper panel. The red line indicates the synthetic curve computed using all the frequencies detected.
\label{fig:LC-timestring-fit}}
\end{figure}

Combination frequencies in $\gamma$ Dor and SPB stars have been widely acknowledged. According to \cite{2015MNRAS.450.3015K}, the complex variability of dozens or hundreds of frequencies can be explained by a few base frequencies and their combination frequencies. We attempt to identify the base frequencies in the Fourier periodogram to generate the predicted frequencies. The frequency spacings are initially calculated to find the tidal splitting and equal separations. The combination frequencies are then determined based on the difference between the detected and predicted frequency, if the difference is below the $\delta f_{LC}$, the signal will be considered as combination frequency. After subtracting all the harmonic frequencies of the binary orbital period, there are 26 identified frequencies. The combination frequencies identified in Table \ref{tab:group} are labeled into four groups, as shown in Figure \ref{fig:Frequencies-coadd}. We finally extracted 9 base frequencies, and 17 combination frequencies, the method used to determine frequency, amplitude, and phase errors was established by \citet{1999DSSN...13...28M}. Figure \ref{fig:LC-timestring-fit} shows the fitted pulsation light curve for the residuals of \emph{Kepler} LC data with all detected frequencies.

\begin{table}
\caption{The group frequencies of V583 Lyr. There are 26 identified frequencies, including 9 base frequencies, and 17 combination frequencies. The zero-point of the time-scale is 2454953.54.}
\begin{center}
\setlength{\tabcolsep}{1.6mm}{
\begin{tabular}{ccccc}\hline\hline
Combination	&	Frequency	&	Amplitude	&	Phase	&	SNR	\\
 	&	($d^{-1}$)	&	(mmag)	&	($rad/2\pi$)	&	 	\\
\hline
\multicolumn{5}{c}{Fg1} \\														
F42=7f22-12f1+4f4	&	1.4513276(299)	&	0.047(4)	&	0.715(13)	&	5.8 	\\
f14=6f5+8f4-24f1	&	1.908725(17)	&	0.083(4)	&	0.189(7)	&	8.6	\\
f24=f1-f2	&	2.122902(19)	&	0.073(4)	&	0.586(8)	&	6.8	\\
f12=f1-f3	&	2.123993(18)	&	0.078(4)	&	0.045(8)	&	7.2	\\
f3	&	2.245163(4)	&	0.328(4)	&	0.180(2)	&	29.3	\\
f2	&	2.246230(2)	&	0.684(4)	&	0.649(1)	&	61.6	\\
f11	&	2.421866(15)	&	0.090(4)	&	0.060(7)	&	7.9	\\
f20=4f11-2f3-f22	&	2.595199(20)	&	0.070(4)	&	0.611(8)	&	6.1	\\
f21=2f11-f3	&	2.598640(20)	&	0.070(4)	&	0.832(8)	&	6	\\
f22	&	2.602494(25)	&	0.057(4)	&	0.420(11)	&	5.8	\\
F27=2f22-2f11+f3	&	2.6071137(226)	&	0.062(4)	&	0.747(10)	&	5.4 	\\
f7=2f19-f5	&	2.781264(8)	&	0.170(4)	&	0.715(3)	&	15.3	\\
f9	&	2.790929(12)	&	0.114(4)	&	0.785(5)	&	10.2	\\
\multicolumn{5}{c}{Fg2} \\									
f18=3f1-f4	&	4.191528(17)	&	0.082(4)	&	0.609(7)	&	8.4	\\
f43=5f1-4f2-3f4	&	4.300665(30)	&	0.047(4)	&	0.174(13)	&	5.6	\\
f19	&	4.344419(19)	&	0.074(4)	&	0.826(8)	&	8	\\
F13=3f9+6f3-f1	&	4.3663826(156)	&	0.09(4)	&	0.128(7)	&	9.6 	\\
f1	&	4.369135(1)	&	1.489(4)	&	0.577(0)	&	160.5	\\
f16=6f1-3f9-6f3	&	4.371668(21)	&	0.067(4)	&	0.704(9)	&	7.3	\\
f15=2f1-f19	&	4.393832(17)	&	0.080(4)	&	0.753(7)	&	8.8	\\
\multicolumn{5}{c}{Fg3} \\									
f5	&	5.906773(72)	&	0.194(4)	&	0.615(3)	&	42.9	\\
f8=2f1-f9	&	5.947306(8)	&	0.164(4)	&	0.247(4)	&	36.7	\\
f47=2f1-2f19+f5	&	5.957014(32)	&	0.043(4)	&	0.218(14)	&	9.6	\\
\multicolumn{5}{c}{Fg4} \\									
f33=2f1	&	8.738301(26)	&	0.054(4)	&	0.084(11)	&	19.3	\\
f4	&	8.915919(5)	&	0.278(4)	&	0.947(2)	&	102.1	\\
f10=15f1+f9-f5-6f4	&	8.940641(13)	&	0.107(4)	&	0.510(6)	&	39.4	\\
\hline
\end{tabular}}
\end{center}
\label{tab:group}
\end{table}

\begin{table}
\caption{Possible rotational splittings. $\delta f$ is the frequency spacing in $d^{-1}$. The $l$ and $m$ values were identified by spacing constraints. The "-" sign is for modes which have not been determined.}
\begin{center}
\setlength{\tabcolsep}{1.6mm}{
\begin{tabular}{cccccc}\hline\hline
Label	 & 	ID	 & 	Frequency 	 & 	$\delta f$	 & 	&  	\\
 	&		&	($d^{-1}$)	&	($d^{-1}$)	& $l$ 	&   $m$	 	\\
\hline
\multirow{2}{*}{1}	 & 	f20	 & 	2.595197 	 & 	\multirow{2}{*}{0.195733}	 & 	1	 & 	0	\\
	 & 	f9	 & 	2.790930 	 & 		 & 	1	 & 	1	\\
	 & 		 & 		 & 		 & 		 & 		\\
\multirow{4}{*}{2}	 & 	f3	 & 	2.245162 	 & 	\multirow{3}{*}{0.176784}	 & 	1	 & 	-1	\\
	 & 	f11	 & 	2.421871 	 & 		 & 	1	 & 	0	\\
	 & 	f21	 & 	2.598655 	 & 		 & 	1	 & 	1	\\
	 & 		 & 		 & 		 & 		 & 		\\
\multirow{2}{*}{3}	 & 	f2	 & 	2.246233 	 & 	\multirow{2}{*}{0.347755}	 & 	2	 & 	0	\\
	 & 	f23	 & 	2.593988 	 & 		 & 	2	 & 	1	\\
	 & 		 & 		 & 		 & 		 & 		\\
\multirow{3}{*}{4}	 & 	f9	 & 	2.790930 	 & 	\multirow{3}{*}{1.578205}	 & 	1	 & 	-	\\
	 & 	f1	 & 	4.369135 	 & 		 & 	1	 & 	-	\\
	 & 	f8	 & 	5.947306 	 & 		 & 		 & 		\\
	 & 		 & 		 & 		 & 		 & 		\\
\multirow{2}{*}{5}	 & 	f9	 & 	2.790930 	 & 	\multirow{2}{*}{1.579225}	 & 	1	 & 	-	\\
	 & 	f44	 & 	4.370156 	 & 		 & 	1	 & 	-	\\
	 & 		 & 		 & 		 & 		 & 		\\
\multirow{2}{*}{6}	 & 	f47	 & 	5.957014 	 & 	\multirow{2}{*}{2.958905}	 & 	2	 & 	-	\\
	 & 	f4	 & 	8.915919 	 & 		 & 	2	 & 	-	\\
\hline
\end{tabular}}
\end{center}
\label{tab:spacing}
\end{table}

\begin{figure}[ht!]
\plotone{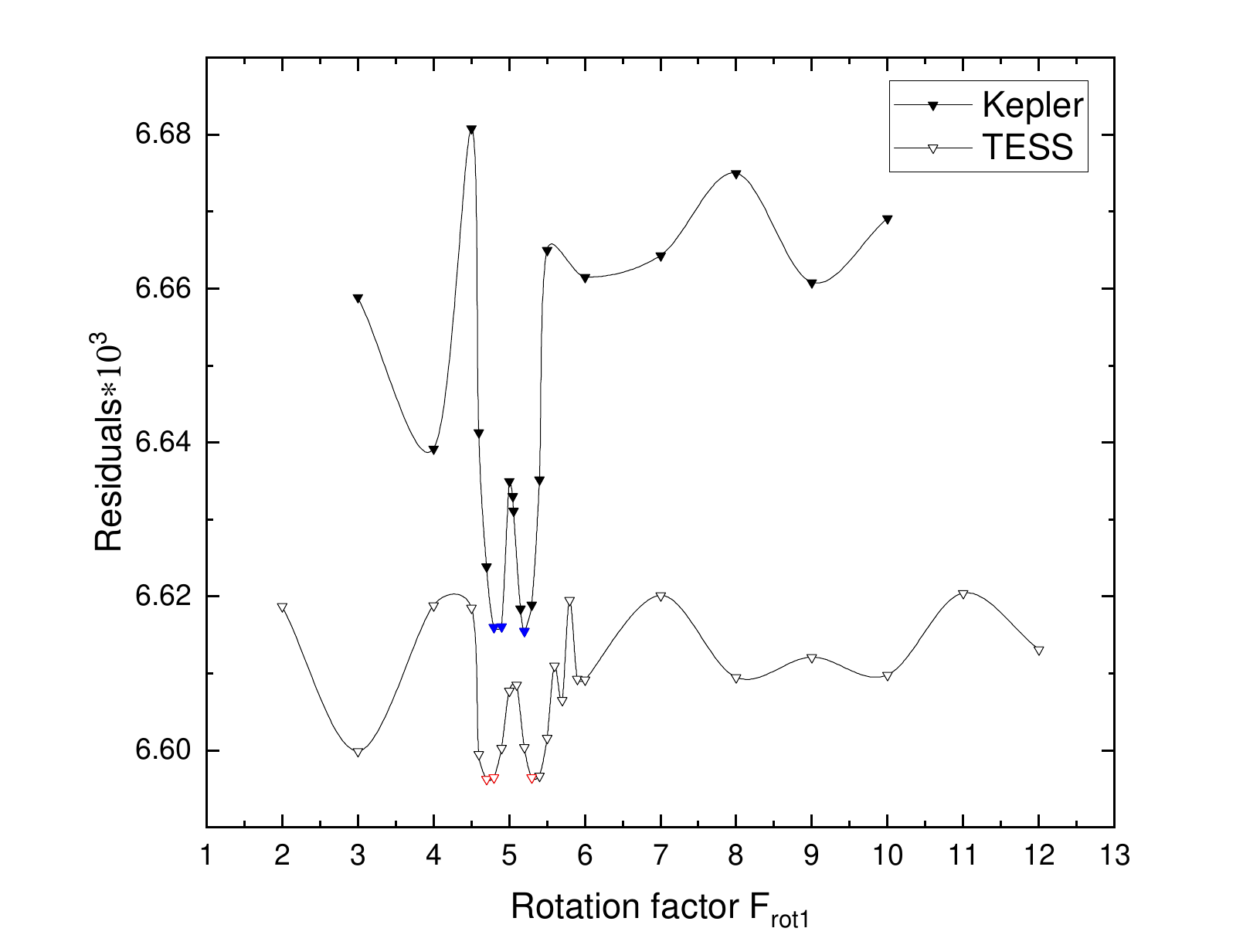}
\caption{Results for a set of assigned gainer rotation parameters $F_{rot1}$ modeled from \emph{TESS} and \emph{Kepler} photometry. All minima fall between 4.7 and 5.3, and the best $F_{rot1}$ = 4.8, implying essentially uniform gainer rotation. 
\label{fig:rotation factor}}
\end{figure}

Mode identification is based on the rotational splitting constraints of the $g$ mode. To ensure certainty in identification, the spacings between and within multiplets must conform to the theory's demands. \cite{1992ApJ...394..670D} derived an approximate expression for the rotational splitting ($\delta f_{l}$) and rotational period ($P_{rot}$) for $g$ modes. The $g$-mode rotational splitting characteristics are used to identify pulsation modes. Table \ref{tab:spacing} displays the $g$-mode rotational splitting for $l$ = 1, $l$ = 2, where $l$ is the angular degree of the oscillation, $m$ is the azimuthal quantum number. The $l$ and $m$ values listed in Table \ref{tab:spacing} were identified by spacing constraints. The observed rotational splitting demonstrates close agreement with theoretical predictions, thereby providing additional support for the validity of the $l$ values. The frequency intervals $\delta f_{l=1}$ $\approx$ 0.1863 $d^{-1}$, $\delta f_{l=2}$ $\approx$ 0.3477 $d^{-1}$ correspond to the g-mode splitting ratio $R_{1,2}$ = $\delta f_{l=1}$ : $\delta f_{l=2}$ = 0.5358, which is within a few percent of its asymptotic value $\frac{1}{\sqrt{3}}$ \citep{1991ApJ...378..326W}. The rotation period of the pulsating component is estimated as $P_{rot1}$ = $\frac{1-[l(l+1)]^{-1}}{\delta f_{l}}$ $\approx$ 2.551 $d$. We have also identified additional sets of frequencies that have the same ratio of frequency spacings. Based on this, we try to speculate on their possible values of $l$, as indicated in Table \ref{tab:spacing}. For further determination of the gainer rotation, we use the photometric method with the W-D model. The rotation of the gainer can be estimated from the photometric solutions of the mean gainer rotation factor $F_{rot1}$ = $\frac{v_{equatorial}}{v_{synchronous}}$, which is defined as the ratio of equatorial and synchronous velocity. For a set of assigned gainer rotation factors $F_{rot1}$, we ran two groups of W-D models on the \emph{TESS} and \emph{Kepler} light curves. Figure \ref{fig:rotation factor} shows the results of Residuals-$F_{rot1}$, the minimum fall between 4.7 and 5.3. The optimal value for $F_{rot1}$ with the least residuals is $F_{rot1}$ = 4.8, supporting nearly uniform gainer rotation in V583 Lyr from accreted angular momentum. The photometric solution is used to compute the $P_{rot1}$ $\approx$ 2.3454 $d$, which is close to the rotation period calculated from the g-mode splitting. The apparent $g$-mode pulsations and the implied A spectral class for the mass-accreting primary suggest that the system may belong to a class of Algols known as oscillating EA or "oEA", such as AS Eri \citep{2004A&A...419.1015M}, for which the primary is a $\delta$ Scuti star. The pulsations could enhance systemic mass loss.

\section{Conclusion and Discussion} \label{sec:Conclusion}

This paper focuses on the detailed analyses of V583 Lyr, an eclipsing binary observed by the \emph{Kepler} mission. New photometric data from \emph{TESS}, together with \emph{Kepler} observations, constrain a very low mass ratio for the system ($q$ = 0.1 $\pm$ 0.004). Our photometric analysis yielded results for the mass ratio $q$, orbital inclination $i$, and the relative parameters listed in Table \ref{tab:wd-results}, which align with the conclusions presented in \cite{Zhang_2020}. The ground-based spectroscopic observations, in conjunction with the results derived from \emph{Gaia}'s BP/RP spectra, permitted the estimation of the hotter star's temperature. According to \cite{1988BAICz..39..329H}, the effective temperature of 9000 K, determined by spectral modeling, may be underestimated if the star has a mass of 3.56 $M_{\odot}$. It is possible that due to the mass transfer the primary star is surrounded by stellar matter or an accretion disk. If a lower $T_{eff}$ is adopted for B-type stars, the $\log$ g values derived from the Balmer line widths will be too small. Consequently, the secondary effective temperature ($T_{eff2}$) and the luminosity of the stars, $L=4 \pi R^{2}\sigma T_{eff}^{4}$, modeled by the W-D program might be underestimated for a given $T_{eff1}$. Future observations and studies on V583 Lyr may allow for a more precise determination of the effective temperature, further confirming the classification of the pulsating primary star. Based on the effective temperature range of SPBs proposed by \cite{2023ApJS..268...16S}, if the pulsating primary has a higher $T_{eff1}$, it could be classified as an SPB star. New spectroscopic observations of V583 Lyr were conducted in specific orbital phases, providing additional data for the construction of the Radial velocity curve. The radial velocity curve of the secondary component has facilitated the most accurate photometric modeling to date, resulting in the derivation of the absolute parameters of both components. Based on our radial velocity analysis, we determined that the orbit is circular and the half amplitude is $K_{2}$ = 135.5 $\pm$ 6.0 km $s^{-1}$. In addition, the system velocity is $\gamma$ = 53.7 $\pm$ 3 km $s^{-1}$. It was found that the mass of the primary component is $M_{1}$ $\approx$ 3.56 $M_{\odot}$, therefore the $M_{2}$ is calculated as $q \cdot M_{2}$ $\approx$ 0.36 $M_{\odot}$.

V583 Lyr is discovered to be an Algol-type eclipsing binary with a gainer surrounded by an accretion disk. We examined the double-peaked emission in $H_{\alpha}$ lines. The variation of the double-peaked emission was explained based on the spectroscopic observations. We also gave the argument of the mass transfer by considering the O-C analysis on the orbital period, from which the matter accretion is estimated to be a rate at $\dot{M_{1}}$ = 3.384 $\times10^{-8}$ $M_{\odot}\cdot yr^{-1}$. The direction of mass transfer in V583 Lyr has been confirmed to be from the low-mass, filled secondary star to the massive primary star via an accretion disk. Based on our determination that the size of the gainer is $R_{1}/a$ $\approx$ 0.074, which falls below the $\omega_{min}$ curve identified by \cite{1975ApJ...198..383L}, a stable accretion disk is confirmed in V583 Lyr. The properties of V583 Lyr are similar to those of the well-studied system TT Hya \citep{1993MNRAS.262..220V,2007ApJ...656.1075M}, whose $R_{1}/a$ = 0.086 and $q$ = 0.226.

We report the first detection of pulsations in V583 Lyr. The primary component of the system was found to be a $g$-mode pulsating star. Pulsation in V583 Lyr is detected in low frequencies, all the 26 detected frequencies are extracted lower than 9 $d^{-1}$. The frequency groups are predominantly composed of a few fundamental frequencies combined together, and we interpret these fundamental frequencies as originating from a small set of $g$ modes. The frequency groups appear in the $g$-mode pulsators such as SPBs and $\gamma$ Dor stars \citep{2012AJ....143..101M,2015MNRAS.450.3015K}. This observational result greatly simplifies the study of g-mode pulsations. Mode identification is based on the rotational splitting properties of the $g$ mode. We have identified the $g$-mode rotational splitting of $l$ = 1, $l$ = 2. Their rotational splitting ratio is in close agreement with theoretical predictions, which reinforces the validity of the $l$ values. The rotation period was estimated as $P_{orb}$ $\approx$ 2.551 $d$, which was also checked by the photometric method of binary modeling. The fitting of the \emph{TESS} and \emph{Kepler} data results in a consistent range of optimal solutions for the $F_{rot1}$ (see Figure \ref{fig:rotation factor}).
The optimal value for $F_{rot1}$ with the least residuals is $F_{rot1}$ = 4.8. This value aligns with the findings obtained from frequency analysis. The differential rotation can be explained by the mass accretion and the non-conservative angular momentum evolution \citep{2010MNRAS.406.1071D}.

\section{Acknowledgments}

The research presented here is supported by the National Key R\&D Program of China (grant No:2022YFE0116800), the National Natural Science Foundation of China (No. 11933008), the Young Talent Project of “Yunnan Revitalization Talent Support Program” in Yunnan Province, and the basic research project of Yunnan Province (Grant No. 202201AT070092). The photometric data used in this paper were obtained from the \emph{Kepler} mission and the \emph{TESS} mission, obtained from the MAST data archive at the Space Telescope Science Institute (STScI). Funding for the \emph{TESS} mission is provided by the NASA Explorer Program. The spectral data were observed by the \emph{LAMOST} telescope and the 2.16 m telescope at Xinglong Station, National Astronomical Observatories. The Guoshoujing Telescope (the Large Sky Area Multi-Object Fiber Spectroscopic Telescope \emph{LAMOST}) is a national major scientific project built by the Chinese Academy of Sciences. The project was funded by the National Development and Reform Commission.\emph{LAMOST} is operated and managed by the National Astronomical Observatories, Chinese Academy of Sciences. Funding for the \emph{Kepler} mission is provided by the NASA Science Mission Directorate. We acknowledge the assistance of the Xinglong 2.16 m Telescope staff. This work was supported in part by the Open Project Program of the Key Laboratory of Optical Astronomy, National Astronomical Observatories, Chinese Academy of Sciences.

%






\appendix 
\section{Appendix information} \label{sec:appendix}
\setcounter{table}{0}
\renewcommand{\thetable}{A\arabic{table}}

All the eclipse times of V583 Lyr are provided in a machine readable format in Table \ref{tab:min}.
\begin{longtable}{p{3.5cm} p{1.5cm}<{\centering} p{1cm}<{\centering} p{1.5cm}<{\centering} p{3.5cm}<{\centering}} 
\caption{All the eclipse times of V583 Lyr. “p/s” means the primary minima or secondary minima.} \\
\label{tab:min}\\
            \toprule 
            Eclipse Times &	Errors	&	p/s	&	Method	&	Observer	\\
            BJD$-2400000$ &	(d)	&	 	&	 	&	 	\\
            \midrule  
                \endfirsthead
                \multicolumn{5}{c}{Eclipse times of V583 Lyr(continue)}\\
                \toprule
                Eclipse Times	&	error	&	p/s	&	Method	&	Observer	\\
                BJD$-2400000$	&	(d)	&	 	&	 	&	 	\\
                \midrule
                \endhead\\
                \bottomrule
                \multicolumn{5}{c}{}
            \endfoot \\
            \bottomrule
            \endlastfoot
26919.59	&	-	&	p	&	pg	&	Guilbaut Peter	\\
28056.617	&	-	&	p	&	pg	&	Guilbaut Peter	\\
29812.64	&	-	&	p	&	pg	&	Guilbaut Peter	\\
30251.547	&	-	&	p	&	pg	&	Guilbaut Peter	\\
35779.286	&	-	&	p	&	pg	&	Berthold Thomas	\\
36792.425	&	-	&	p	&	pg	&	Berthold Thomas	\\
36837.357	&	-	&	p	&	pg	&	Berthold Thomas	\\
37028.617	&	-	&	p	&	pg	&	Berthold Thomas	\\
37659.23	&	-	&	p	&	pg	&	Berthold Thomas	\\
38503.584	&	-	&	p	&	pg	&	Berthold Thomas	\\
39021.359	&	-	&	p	&	pg	&	Berthold Thomas	\\
39055.319	&	-	&	p	&	pg	&	Berthold Thomas	\\
40023.462	&	-	&	p	&	pg	&	Berthold Thomas	\\
40383.468	&	-	&	p	&	pg	&	Berthold Thomas	\\
40507.326	&	-	&	p	&	pg	&	Berthold Thomas	\\
41982.314	&	-	&	p	&	pg	&	Berthold Thomas	\\
43659.742	&	-	&	p	&	pg	&	Guilbaut Peter	\\
43670.744	&	-	&	p	&	pg	&	Guilbaut Peter	\\
44346.554	&	-	&	p	&	pg	&	Berthold Thomas	\\
45089.755	&	-	&	p	&	pg	&	Guilbaut Peter	\\
45494.698	&	-	&	p	&	pg	&	Guilbaut Peter	\\
45618.49	&	-	&	p	&	pg	&	Guilbaut Peter	\\
45911.425	&	-	&	p	&	pg	&	Berthold Thomas	\\
45990.285	&	-	&	p	&	pg	&	Berthold Thomas	\\
46001.271	&	-	&	p	&	pg	&	Berthold Thomas	\\
46237.742	&	-	&	p	&	pg	&	Guilbaut Peter	\\
46260.459	&	-	&	p	&	pg	&	Berthold Thomas	\\
46316.628	&	-	&	p	&	pg	&	Guilbaut Peter	\\
46699.545	&	-	&	p	&	pg	&	Guilbaut Peter	\\
46733.468	&	-	&	p	&	pg	&	Guilbaut Peter	\\
48095.44	&	-	&	p	&	pg	&	Berthold Thomas	\\
50279.445	&	-	&	p	&	pg	&	Berthold Thomas	\\
51101.31	&	-	&	p	&	pg	&	Dahlmark L	\\
51275.424	&	-	&	s	&	ccd	&	Paschke Anton	\\
51270.02	&	-	&	p	&	-	&	ROTSE	\\
53262.62	&	-	&	p	&	-	&	Krajci estimate	\\
53296.4	&	-	&	p	&	-	&	Krajci estimate	\\
54962.54906	&	0.00118	&	p	&	ccd	&	Kepler	\\
54973.80475	&	0.00495	&	p	&	ccd	&	Kepler	\\
54985.06057	&	0.00503	&	p	&	ccd	&	Kepler	\\
54996.32747	&	0.00398	&	p	&	ccd	&	Kepler	\\
55007.58176	&	0.00346	&	p	&	ccd	&	Kepler	\\
55018.83729	&	0.00216	&	p	&	ccd	&	Kepler	\\
55030.09478	&	0.00224	&	p	&	ccd	&	Kepler	\\
55041.34874	&	0.00305	&	p	&	ccd	&	Kepler	\\
55052.60903	&	0.00176	&	p	&	ccd	&	Kepler	\\
55075.12352	&	0.00109	&	p	&	ccd	&	Kepler	\\
55086.37657	&	0.00091	&	p	&	ccd	&	Kepler	\\
55097.63569	&	0.00102	&	p	&	ccd	&	Kepler	\\
55108.8938	&	0.00084	&	p	&	ccd	&	Kepler	\\
55120.15145	&	0.00131	&	p	&	ccd	&	Kepler	\\
55131.40105	&	0.00113	&	p	&	ccd	&	Kepler	\\
55142.66566	&	0.00117	&	p	&	ccd	&	Kepler	\\
55153.92044	&	0.00053	&	p	&	ccd	&	Kepler	\\
55165.1765	&	0.00077	&	p	&	ccd	&	Kepler	\\
55176.4326	&	0.00177	&	p	&	ccd	&	Kepler	\\
55187.69212	&	0.00109	&	p	&	ccd	&	Kepler	\\
55198.95328	&	0.00073	&	p	&	ccd	&	Kepler	\\
55210.21672	&	0.00114	&	p	&	ccd	&	Kepler	\\
55221.46412	&	0.001	&	p	&	ccd	&	Kepler	\\
55243.98384	&	0.00114	&	p	&	ccd	&	Kepler	\\
55255.25143	&	0.00162	&	p	&	ccd	&	Kepler	\\
55266.50145	&	0.00106	&	p	&	ccd	&	Kepler	\\
55277.76019	&	0.00128	&	p	&	ccd	&	Kepler	\\
55289.01458	&	0.00101	&	p	&	ccd	&	Kepler	\\
55300.27322	&	0.00099	&	p	&	ccd	&	Kepler	\\
55311.53429	&	0.0009	&	p	&	ccd	&	Kepler	\\
55322.79061	&	0.00115	&	p	&	ccd	&	Kepler	\\
55334.052	&	0.00092	&	p	&	ccd	&	Kepler	\\
55345.30826	&	0.00079	&	p	&	ccd	&	Kepler	\\
55356.56225	&	0.00059	&	p	&	ccd	&	Kepler	\\
55367.82196	&	0.00152	&	p	&	ccd	&	Kepler	\\
55379.07842	&	0.00252	&	p	&	ccd	&	Kepler	\\
55390.33124	&	0.00058	&	p	&	ccd	&	Kepler	\\
55401.58704	&	0.0017	&	p	&	ccd	&	Kepler	\\
55412.84894	&	0.00117	&	p	&	ccd	&	Kepler	\\
55424.10409	&	0.00154	&	p	&	ccd	&	Kepler	\\
55435.36015	&	0.00095	&	p	&	ccd	&	Kepler	\\
55446.62026	&	0.00119	&	p	&	ccd	&	Kepler	\\
55457.88401	&	0.00063	&	p	&	ccd	&	Kepler	\\
55469.13395	&	0.00118	&	p	&	ccd	&	Kepler	\\
55480.39487	&	0.00066	&	p	&	ccd	&	Kepler	\\
55491.64639	&	0.00106	&	p	&	ccd	&	Kepler	\\
55502.91536	&	0.00077	&	p	&	ccd	&	Kepler	\\
55514.17041	&	0.00089	&	p	&	ccd	&	Kepler	\\
55525.4341	&	0.00077	&	p	&	ccd	&	Kepler	\\
55536.68032	&	0.00092	&	p	&	ccd	&	Kepler	\\
55547.93501	&	0.00098	&	p	&	ccd	&	Kepler	\\
55570.45358	&	0.00072	&	p	&	ccd	&	Kepler	\\
55581.7068	&	0.00068	&	p	&	ccd	&	Kepler	\\
55592.97095	&	0.00155	&	p	&	ccd	&	Kepler	\\
55604.22349	&	0.00069	&	p	&	ccd	&	Kepler	\\
55615.47533	&	0.00109	&	p	&	ccd	&	Kepler	\\
55626.74445	&	0.00092	&	p	&	ccd	&	Kepler	\\
55649.26482	&	0.00105	&	p	&	ccd	&	Kepler	\\
55660.51934	&	0.00147	&	p	&	ccd	&	Kepler	\\
55671.78466	&	0.00132	&	p	&	ccd	&	Kepler	\\
55683.04284	&	0.00104	&	p	&	ccd	&	Kepler	\\
55694.29844	&	0.00074	&	p	&	ccd	&	Kepler	\\
55705.55047	&	0.001	&	p	&	ccd	&	Kepler	\\
55716.80923	&	0.00081	&	p	&	ccd	&	Kepler	\\
55728.06454	&	0.0012	&	p	&	ccd	&	Kepler	\\
55750.58122	&	0.00065	&	p	&	ccd	&	Kepler	\\
55761.84343	&	0.00087	&	p	&	ccd	&	Kepler	\\
55773.09822	&	0.00074	&	p	&	ccd	&	Kepler	\\
55784.35277	&	0.00072	&	p	&	ccd	&	Kepler	\\
55795.60865	&	0.00075	&	p	&	ccd	&	Kepler	\\
55806.86645	&	0.00236	&	p	&	ccd	&	Kepler	\\
55818.13032	&	0.00082	&	p	&	ccd	&	Kepler	\\
55829.39502	&	0.00121	&	p	&	ccd	&	Kepler	\\
55840.64421	&	0.00122	&	p	&	ccd	&	Kepler	\\
55851.90991	&	0.00188	&	p	&	ccd	&	Kepler	\\
55863.16541	&	0.00061	&	p	&	ccd	&	Kepler	\\
55874.42462	&	0.00103	&	p	&	ccd	&	Kepler	\\
55885.68001	&	0.00071	&	p	&	ccd	&	Kepler	\\
55908.19435	&	0.00057	&	p	&	ccd	&	Kepler	\\
55919.46203	&	0.00148	&	p	&	ccd	&	Kepler	\\
55930.72482	&	0.0012	&	p	&	ccd	&	Kepler	\\
55941.97944	&	0.00058	&	p	&	ccd	&	Kepler	\\
55953.23899	&	0.00097	&	p	&	ccd	&	Kepler	\\
55964.49227	&	0.00105	&	p	&	ccd	&	Kepler	\\
55975.74501	&	0.00084	&	p	&	ccd	&	Kepler	\\
55998.26247	&	0.00083	&	p	&	ccd	&	Kepler	\\
56009.52291	&	0.00096	&	p	&	ccd	&	Kepler	\\
56020.78218	&	0.00113	&	p	&	ccd	&	Kepler	\\
56032.03649	&	0.0004	&	p	&	ccd	&	Kepler	\\
56043.29263	&	0.0017	&	p	&	ccd	&	Kepler	\\
56054.54715	&	0.00095	&	p	&	ccd	&	Kepler	\\
56065.80598	&	0.00192	&	p	&	ccd	&	Kepler	\\
56077.06137	&	0.00097	&	p	&	ccd	&	Kepler	\\
56088.32463	&	0.00085	&	p	&	ccd	&	Kepler	\\
56099.57735	&	0.00102	&	p	&	ccd	&	Kepler	\\
56110.83719	&	0.00112	&	p	&	ccd	&	Kepler	\\
56122.09666	&	0.00096	&	p	&	ccd	&	Kepler	\\
56133.34666	&	0.00062	&	p	&	ccd	&	Kepler	\\
56144.60422	&	0.00096	&	p	&	ccd	&	Kepler	\\
56155.86245	&	0.00471	&	p	&	ccd	&	Kepler	\\
56167.11712	&	0.00063	&	p	&	ccd	&	Kepler	\\
56178.37765	&	0.00087	&	p	&	ccd	&	Kepler	\\
56189.63674	&	0.00114	&	p	&	ccd	&	Kepler	\\
56200.89507	&	0.00043	&	p	&	ccd	&	Kepler	\\
56212.15494	&	0.00137	&	p	&	ccd	&	Kepler	\\
56223.40392	&	0.00059	&	p	&	ccd	&	Kepler	\\
56234.67182	&	0.0007	&	p	&	ccd	&	Kepler	\\
56245.79806	&	0.00968	&	p	&	ccd	&	Kepler	\\
56257.18703	&	0.00069	&	p	&	ccd	&	Kepler	\\
56279.69705	&	0.00116	&	p	&	ccd	&	Kepler	\\
56290.96701	&	0.00087	&	p	&	ccd	&	Kepler	\\
56302.2212	&	0.00109	&	p	&	ccd	&	Kepler	\\
56324.74697	&	0.00132	&	p	&	ccd	&	Kepler	\\
56335.99261	&	0.00128	&	p	&	ccd	&	Kepler	\\
56347.26161	&	0.00143	&	p	&	ccd	&	Kepler	\\
56369.77672	&	0.00121	&	p	&	ccd	&	Kepler	\\
56381.02833	&	0.00079	&	p	&	ccd	&	Kepler	\\
56392.36089	&	0.00757	&	p	&	ccd	&	Kepler	\\
56403.54591	&	0.00132	&	p	&	ccd	&	Kepler	\\
58688.88345	&	0.00103	&	p	&	ccd	&	TESS	\\
58700.14082	&	0.00103	&	p	&	ccd	&	TESS	\\
59398.12574	&	0.001	&	p	&	ccd	&	TESS	\\
59409.38417	&	0.00099	&	p	&	ccd	&	TESS	\\
59420.64174	&	0.00101	&	p	&	ccd	&	TESS	\\
59431.90059	&	0.00101	&	p	&	ccd	&	TESS	\\
59443.1592	&	0.00101	&	p	&	ccd	&	TESS	\\
59747.12617	&	0.00099	&	p	&	ccd	&	TESS	\\
59758.38281	&	0.00097	&	p	&	ccd	&	TESS	\\
59780.90062	&	0.00101	&	p	&	ccd	&	TESS	\\
59792.15818	&	0.00098	&	p	&	ccd	&	TESS	\\
54956.90505	&	0.00281	&	s	&	ccd	&	Kepler	\\
54968.16029	&	0.00203	&	s	&	ccd	&	Kepler	\\
54979.4222	&	0.00226	&	s	&	ccd	&	Kepler	\\
54990.68071	&	0.00284	&	s	&	ccd	&	Kepler	\\
55013.19861	&	0.0025	&	s	&	ccd	&	Kepler	\\
55024.46451	&	0.00175	&	s	&	ccd	&	Kepler	\\
55035.72404	&	0.00226	&	s	&	ccd	&	Kepler	\\
55046.97638	&	0.00348	&	s	&	ccd	&	Kepler	\\
55058.23075	&	0.00284	&	s	&	ccd	&	Kepler	\\
55069.497	&	0.00201	&	s	&	ccd	&	Kepler	\\
55080.75517	&	0.00334	&	s	&	ccd	&	Kepler	\\
55103.27367	&	0.00235	&	s	&	ccd	&	Kepler	\\
55114.51521	&	0.00833	&	s	&	ccd	&	Kepler	\\
55125.77682	&	0.00335	&	s	&	ccd	&	Kepler	\\
55137.03481	&	0.00261	&	s	&	ccd	&	Kepler	\\
55148.30066	&	0.0025	&	s	&	ccd	&	Kepler	\\
55159.56326	&	0.00236	&	s	&	ccd	&	Kepler	\\
55170.80649	&	0.0032	&	s	&	ccd	&	Kepler	\\
55182.08216	&	0.00297	&	s	&	ccd	&	Kepler	\\
55193.33279	&	0.0032	&	s	&	ccd	&	Kepler	\\
55204.58318	&	0.00368	&	s	&	ccd	&	Kepler	\\
55215.83515	&	0.00245	&	s	&	ccd	&	Kepler	\\
55227.10101	&	0.00183	&	s	&	ccd	&	Kepler	\\
55238.36215	&	0.00416	&	s	&	ccd	&	Kepler	\\
55249.61753	&	0.00366	&	s	&	ccd	&	Kepler	\\
55260.85453	&	0.00252	&	s	&	ccd	&	Kepler	\\
55272.13018	&	0.00282	&	s	&	ccd	&	Kepler	\\
55283.36942	&	0.00412	&	s	&	ccd	&	Kepler	\\
55294.63165	&	0.00194	&	s	&	ccd	&	Kepler	\\
55305.8954	&	0.00234	&	s	&	ccd	&	Kepler	\\
55317.13496	&	0.00419	&	s	&	ccd	&	Kepler	\\
55328.38962	&	0.00337	&	s	&	ccd	&	Kepler	\\
55339.66594	&	0.00267	&	s	&	ccd	&	Kepler	\\
55350.91955	&	0.00323	&	s	&	ccd	&	Kepler	\\
55362.17814	&	0.00303	&	s	&	ccd	&	Kepler	\\
55373.43837	&	0.00264	&	s	&	ccd	&	Kepler	\\
55384.70656	&	0.00191	&	s	&	ccd	&	Kepler	\\
55395.9708	&	0.00228	&	s	&	ccd	&	Kepler	\\
55407.20812	&	0.00292	&	s	&	ccd	&	Kepler	\\
55418.49212	&	0.00245	&	s	&	ccd	&	Kepler	\\
55429.73445	&	0.00303	&	s	&	ccd	&	Kepler	\\
55441.00253	&	0.0042	&	s	&	ccd	&	Kepler	\\
55452.24059	&	0.00297	&	s	&	ccd	&	Kepler	\\
55463.51822	&	0.00323	&	s	&	ccd	&	Kepler	\\
55474.77325	&	0.00434	&	s	&	ccd	&	Kepler	\\
55486.0278	&	0.00271	&	s	&	ccd	&	Kepler	\\
55497.28764	&	0.00211	&	s	&	ccd	&	Kepler	\\
55508.54619	&	0.00325	&	s	&	ccd	&	Kepler	\\
55519.79544	&	0.00309	&	s	&	ccd	&	Kepler	\\
55531.05527	&	0.00241	&	s	&	ccd	&	Kepler	\\
55542.31538	&	0.00276	&	s	&	ccd	&	Kepler	\\
55576.09074	&	0.00239	&	s	&	ccd	&	Kepler	\\
55587.34936	&	0.00261	&	s	&	ccd	&	Kepler	\\
55598.60432	&	0.00391	&	s	&	ccd	&	Kepler	\\
55609.8563	&	0.00255	&	s	&	ccd	&	Kepler	\\
55621.11857	&	0.00226	&	s	&	ccd	&	Kepler	\\
55632.37231	&	0.00291	&	s	&	ccd	&	Kepler	\\
55643.61911	&	0.003	&	s	&	ccd	&	Kepler	\\
55654.88322	&	0.00233	&	s	&	ccd	&	Kepler	\\
55666.14079	&	0.00353	&	s	&	ccd	&	Kepler	\\
55677.40812	&	0.00312	&	s	&	ccd	&	Kepler	\\
55688.65996	&	0.00283	&	s	&	ccd	&	Kepler	\\
55699.92197	&	0.00268	&	s	&	ccd	&	Kepler	\\
55711.17194	&	0.003	&	s	&	ccd	&	Kepler	\\
55722.42968	&	0.00292	&	s	&	ccd	&	Kepler	\\
55733.69563	&	0.00298	&	s	&	ccd	&	Kepler	\\
55744.96258	&	0.00268	&	s	&	ccd	&	Kepler	\\
55756.21379	&	0.00245	&	s	&	ccd	&	Kepler	\\
55767.4725	&	0.00238	&	s	&	ccd	&	Kepler	\\
55778.72901	&	0.00301	&	s	&	ccd	&	Kepler	\\
55789.98065	&	0.00268	&	s	&	ccd	&	Kepler	\\
55801.23973	&	0.00263	&	s	&	ccd	&	Kepler	\\
55812.49611	&	0.003	&	s	&	ccd	&	Kepler	\\
55823.75901	&	0.0029	&	s	&	ccd	&	Kepler	\\
55835.0063	&	0.00223	&	s	&	ccd	&	Kepler	\\
55846.26431	&	0.00238	&	s	&	ccd	&	Kepler	\\
55857.51895	&	0.00312	&	s	&	ccd	&	Kepler	\\
55868.78511	&	0.00308	&	s	&	ccd	&	Kepler	\\
55880.05026	&	0.0027	&	s	&	ccd	&	Kepler	\\
55891.31539	&	0.00266	&	s	&	ccd	&	Kepler	\\
55902.57346	&	0.00372	&	s	&	ccd	&	Kepler	\\
55913.82827	&	0.003	&	s	&	ccd	&	Kepler	\\
55925.09629	&	0.00238	&	s	&	ccd	&	Kepler	\\
55936.34373	&	0.0037	&	s	&	ccd	&	Kepler	\\
55947.59831	&	0.00362	&	s	&	ccd	&	Kepler	\\
55958.73708	&	0.00249	&	s	&	ccd	&	Kepler	\\
55970.11253	&	0.0035	&	s	&	ccd	&	Kepler	\\
55981.37338	&	0.00408	&	s	&	ccd	&	Kepler	\\
55992.62052	&	0.00324	&	s	&	ccd	&	Kepler	\\
56003.88844	&	0.00254	&	s	&	ccd	&	Kepler	\\
56026.39139	&	0.00399	&	s	&	ccd	&	Kepler	\\
56037.6573	&	0.00242	&	s	&	ccd	&	Kepler	\\
56048.91687	&	0.00539	&	s	&	ccd	&	Kepler	\\
56060.17711	&	0.00342	&	s	&	ccd	&	Kepler	\\
56071.42991	&	0.00329	&	s	&	ccd	&	Kepler	\\
56082.69783	&	0.00263	&	s	&	ccd	&	Kepler	\\
56093.95032	&	0.0034	&	s	&	ccd	&	Kepler	\\
56105.21256	&	0.00465	&	s	&	ccd	&	Kepler	\\
56116.46626	&	0.00242	&	s	&	ccd	&	Kepler	\\
56150.23627	&	0.00379	&	s	&	ccd	&	Kepler	\\
56161.49621	&	0.00263	&	s	&	ccd	&	Kepler	\\
56172.75896	&	0.00356	&	s	&	ccd	&	Kepler	\\
56184.01362	&	0.00292	&	s	&	ccd	&	Kepler	\\
56195.27489	&	0.00318	&	s	&	ccd	&	Kepler	\\
56217.78816	&	0.00444	&	s	&	ccd	&	Kepler	\\
56229.03941	&	0.00339	&	s	&	ccd	&	Kepler	\\
56240.31089	&	0.00244	&	s	&	ccd	&	Kepler	\\
56251.59865	&	0.00696	&	s	&	ccd	&	Kepler	\\
56262.82947	&	0.00344	&	s	&	ccd	&	Kepler	\\
56274.0871	&	0.00258	&	s	&	ccd	&	Kepler	\\
56285.34141	&	0.00259	&	s	&	ccd	&	Kepler	\\
56296.60346	&	0.00298	&	s	&	ccd	&	Kepler	\\
56307.84644	&	0.00249	&	s	&	ccd	&	Kepler	\\
56330.37344	&	0.00385	&	s	&	ccd	&	Kepler	\\
56341.61468	&	0.00425	&	s	&	ccd	&	Kepler	\\
56352.87622	&	0.00254	&	s	&	ccd	&	Kepler	\\
56364.1289	&	0.00308	&	s	&	ccd	&	Kepler	\\
56375.39187	&	0.00393	&	s	&	ccd	&	Kepler	\\
56386.64076	&	0.00329	&	s	&	ccd	&	Kepler	\\
56397.91581	&	0.00168	&	s	&	ccd	&	Kepler	\\
56409.16968	&	0.00302	&	s	&	ccd	&	Kepler	\\
56420.42958	&	0.00367	&	s	&	ccd	&	Kepler	\\
58694.50683	&	0.00128	&	s	&	ccd	&	TESS	\\
58705.76915	&	0.00119	&	s	&	ccd	&	TESS	\\
59392.50503	&	0.00116	&	s	&	ccd	&	TESS	\\
59403.77771	&	0.00112	&	s	&	ccd	&	TESS	\\
59415.02322	&	0.00086	&	s	&	ccd	&	TESS	\\
59426.27554	&	0.001	&	s	&	ccd	&	TESS	\\
59437.53455	&	0.00114	&	s	&	ccd	&	TESS	\\
59752.76346	&	0.00109	&	s	&	ccd	&	TESS	\\
59764.02365	&	0.00096	&	s	&	ccd	&	TESS	\\
59775.27987	&	0.00109	&	s	&	ccd	&	TESS	\\
59786.53376	&	0.00116	&	s	&	ccd	&	TESS	\\
        \end{longtable}

\clearpage\newpage


\bibliography{sample631}{}
\bibliographystyle{aasjournal}



\end{document}